\documentstyle{mn}
\input epsf

\newif\ifAMStwofonts

\ifoldfss
  \ifCUPmtlplainloaded \else
    \NewTextAlphabet{textbfit} {cmbxti10} {}
    \NewTextAlphabet{textbfss} {cmssbx10} {}
    \NewMathAlphabet{mathbfit} {cmbxti10} {} 
    \NewMathAlphabet{mathbfss} {cmssbx10} {} 
  \fi
  \ifAMStwofonts
    \ifCUPmtlplainloaded \else
      \NewSymbolFont{upmath} {eurm10}
      \NewSymbolFont{AMSa} {msam10}
      \NewMathSymbol{\upi}     {0}{upmath}{19}
      \NewMathSymbol{\umu}     {0}{upmath}{16}
      \NewMathSymbol{\upartial}{0}{upmath}{40}
      \NewMathSymbol{\leqslant}{3}{AMSa}{36}
      \NewMathSymbol{\geqslant}{3}{AMSa}{3E}

       \let\le=\leqslant
       \let\ge=\geqslant
    \fi
  \fi
\fi 

\ifnfssone
  \newmathalphabet{\mathit}
  \addtoversion{normal}{\mathit}{cmr}{m}{it}
  \addtoversion{bold}{\mathit}{cmr}{bx}{it}
  \newmathalphabet{\mathbfit} 
  \addtoversion{normal}{\mathbfit}{cmr}{bx}{it}
  \addtoversion{bold}{\mathbfit}{cmr}{bx}{it}
  \newmathalphabet{\mathbfss} 
  \addtoversion{normal}{\mathbfss}{cmss}{bx}{n}
  \addtoversion{bold}{\mathbfss}{cmss}{bx}{n}
  \ifAMStwofonts
    \ifCUPmtlplainloaded \else
      \UseAMStwoboldmath
      \makeatletter
      \new@mathgroup\upmath@group
      \define@mathgroup\mv@normal\upmath@group{eur}{m}{n}
      \define@mathgroup\mv@bold\upmath@group{eur}{b}{n}
      \edef\UPM{\hexnumber\upmath@group}
      \new@mathgroup\amsa@group
      \define@mathgroup\mv@normal\amsa@group{msa}{m}{n}
      \define@mathgroup\mv@bold\amsa@group{msa}{m}{n}
      \edef\AMSa{\hexnumber\amsa@group}
      \makeatother
      \mathchardef\upi="0\UPM19
      \mathchardef\umu="0\UPM16
      \mathchardef\upartial="0\UPM40
      \mathchardef\leqslant="3\AMSa36
      \mathchardef\geqslant="3\AMSa3E

       \let\le=\leqslant
       \let\ge=\geqslant
    \fi
  \fi
\fi 

\ifnfsstwo
  \DeclareMathAlphabet{\mathbfit}{OT1}{cmr}{bx}{it}
  \SetMathAlphabet\mathbfit{bold}{OT1}{cmr}{bx}{it}
  \DeclareMathAlphabet{\mathbfss}{OT1}{cmss}{bx}{n}
  \SetMathAlphabet\mathbfss{bold}{OT1}{cmss}{bx}{n}
  \ifAMStwofonts
    \ifCUPmtlplainloaded \else
      \DeclareSymbolFont{UPM}{U}{eur}{m}{n}
      \SetSymbolFont{UPM}{bold}{U}{eur}{b}{n}
      \DeclareSymbolFont{AMSa}{U}{msa}{m}{n}
      \DeclareMathSymbol{\upi}{0}{UPM}{"19}
      \DeclareMathSymbol{\umu}{0}{UPM}{"16}
      \DeclareMathSymbol{\upartial}{0}{UPM}{"40}
      \DeclareMathSymbol{\leqslant}{3}{AMSa}{"36}
      \DeclareMathSymbol{\geqslant}{3}{AMSa}{"3E}

       \let\le=\leqslant
       \let\ge=\geqslant
    \fi
  \fi
\fi 

\ifCUPmtlplainloaded \else
  \ifAMStwofonts \else 
    \def\upi{\pi}
    \def\umu{\mu}
    \def\upartial{\partial}
  \fi
\fi

\title[NaSt1: A WR star cloaked by an $\eta$ Car--like nebula?]
{NaSt1: A Wolf-Rayet star cloaked by an $\boldeta$ Car--like nebula?}
\author[P. A. Crowther \& L. J. Smith]
       {Paul A. Crowther and Linda J. Smith \\
        Department of Physics and Astronomy, University College London, Gower
Street, London, WC1E 6BT}
\date{Accepted. Received}

\pagerange{\pageref{firstpage}--\pageref{lastpage}}
\pubyear{1999}

\begin{document}

\maketitle

\label{firstpage}

\begin{abstract}
We present a study of the peculiar Galactic emission line object NaSt1 
(WR122, IRAS 18497+0056) which has previously been classified as a
Wolf-Rayet (WR) star. Our spectroscopic dataset comprises 
Keck~I-HIRES,  WHT-ISIS and UKIRT-CGS4 observations which show that
NaSt1 has a highly reddened nebular spectrum with extremely strong
permitted and forbidden lines covering a wide range  in excitation
(H\,{\sc i}, He\,{\sc i-ii},  N\,{\sc i-iii}, [N\,{\sc ii}], [Ne\,{\sc
iii-iv}], Mg\,{\sc i-ii}, Si\,{\sc ii},
[S\,{\sc ii-iii}], [Ar\,{\sc iii-v}],  [Ca\,{\sc v-vii}], [Fe\,{\sc
ii-vii}], [Ni\,{\sc ii-iii}]). 
[O\,{\sc ii-iii}] is unusually weak, with He\,{\sc i-ii} and [N\,{\sc ii}]
very strong, and carbon absent, suggestive of chemical peculiarities.
Narrow-band WHT imaging reveals an elliptical nebula with an average 
diameter of 6.8 \,arcsec. We measure an interstellar extinction of  $E_{\rm
B-V}$$\sim$2.1\,mag and estimate a distance of 1--3\,kpc,  suggesting
that NaSt1 is a luminous object, with 4$\le$log\,($L/L_{\odot}$)$\le$6.5.
We determine the physical parameters and abundances from the nebular
forbidden lines. For $T_{e}=13\,000$\,K and
$N_{e}=3.10^{6}$\,cm$^{-3}$, we obtain He/H$>0.64$, N enhanced by a
factor of 20, O deficient by a factor of 140, while Ne, Ar and S are
normal compared to average H\,{\sc ii} region abundances. This unusual
abundance pattern suggests that the nebula consists of fully
CNO-processed material. We compare the spectral appearance of NaSt1
with other luminous emission objects, and conclude that it is not an
Ofpe/WN9, B[e] star or symbiotic nova although it does share several
characteristics of these systems. We suggest instead that NaSt1 contains
a massive evolved star that ejected its heavily CNO-processed outer
layers a few thousand years ago. Although the stellar remnant is
completely hidden from view by the dense nebula, we argue that
the star must be an early-type WR star. The only object that 
shares some of the peculiarities of NaSt1 is $\eta$ Carinae.
Whatever its true nature, NaSt1 should no longer 
be considered as a late-WN classification standard in the near-IR.
\end{abstract}

\begin{keywords}
stars: individual (NaSt1) -- stars: emission-line -- stars: peculiar --
stars: Wolf-Rayet -- ISM: H\,{\sc ii} regions
\end{keywords}

\section{Introduction}

Relatively few peculiar emission line objects identified from
H$\alpha$ surveys in the 1960's (e.g. Henize 1967, 1976) have been
studied in detail, principally because they are faint and suffer from
heavy reddening. However, advances with instruments combined with the 
availability of 8--10m telescopes now permit the routine 
observation of such objects, which may provide new information 
on stellar systems and evolution.

One such object is NaSt1 (V$=14.5$\,mag), 
discovered by Nassau \& Stephenson (1963)
who proposed a Wolf-Rayet (WR) classification because of its strong 
emission line spectrum. Its appearance, however, was quite unlike any 
previously known WR star in the Galaxy or Large Magellanic Cloud 
(Massey \& Conti 1983). Nevertheless, Massey \& Conti proposed a cool, 
late type nitrogen sequence WN10 spectral type for NaSt1, and it 
was included in the sixth WR catalogue  as WR122 (van der Hucht et al. 1981).

\begin{table}
\centering
\begin{minipage}{80mm}
\caption{Journal of optical and infrared observations of NaSt1
($\alpha =$ 18\,h 49\,m 44.8 s; $\delta = 00^{\circ}\,56'\,03''$, B1950.0).}
\label{TABLE1}
\begin{tabular}{@{}l@{\hspace{2mm}}l@{\hspace{2mm}}c@{\hspace{1mm}}r
@{\hspace{3mm}}l@{\hspace{1mm}}r}
\hline
\multicolumn{5}{c}{-- Spectroscopy --} \\
Date & Telescope$+$  & Wavelength & Exp. & Sp.\,Res. & PA\\
     & Instrument & (nm)        &(sec)& (nm) & ($^\circ$)\\
\hline
29 Jul 1994 & WHT--ISIS & 382--462     & 1800     & 0.16  & 26\\ 
            &           & 531--622     & 1800     & 0.17  & 26 \\
            &           & 611--703     &   60     & 0.17  & 26 \\
01 Aug 1994 & WHT--ISIS & 382--462     & 1800     & 0.16 & 353 \\
            &           & 458--538     & 1800     & 0.16 & 353 \\
19 Aug 1994 & UKIRT--CGS4 & 1028--1132 &  128     & 1.4  & 90\\
            &             & 2007--2215 &   64     & 2.6  & 90\\
21 Aug 1994 & UKIRT--CGS4 & 1228--1332 &  320     & 1.5  & 90\\
            &             & 1606--1814 &  128     & 2.8  & 90\\
            &             & 2055--2062 & 1280 & 0.13     & 90\\
13 Oct 1994 & Keck~I--HIRES& 435--679    & 1500     & 0.01 & 0\\ 
            &            & 632--875   &  600     & 0.02 & 0\\ 
\hline
\multicolumn{5}{c}{-- Imaging -- } \\
Date & Telescope$+$  & Wavelength & Exp. & FWHM\\
     & Instrument    & (nm)       & (sec) & (nm) \\
\hline
12 Aug 1996 & WHT--AUX  & 469.1     &  350   & 4.9\\  
            &           & 588.1     &  50      & 4.6\\  
            &           & 656.5     &  50      & 6.0\\  %
            &           & 659.0     &  200   & 1.6\\  %
\hline
\end{tabular}
\end{minipage}
\end{table}

More recently, van der Hucht, Williams \& Th\'{e} (1984) and 
Williams, van der Hucht \& Th\'{e} (1987) have obtained infra-red 
(IR) photometry of NaSt1, and the 
closely related object LS4005 (WR85a), revealing the presence of 
circumstellar dust shells. van der Hucht et al. (1989, 1997) reported
moderate IR photometric variability for NaSt1 and argued against a 
WR nature, instead preferring an alternative
(massive) emission line nature (either B[e], O[e] or Ofpe/WN9)
based on optical and infrared spectroscopy. B[e] supergiants appear to 
represent objects with an equatorial excretion disk plus an 
OB-type stellar wind in the polar regions. Ofpe/WN9 stars --
now revised to WN9--11 (Smith, Crowther \& Prinja 1994) --
are intimately related to Luminous Blue Variables (LBVs)
and classical WR stars (Crowther  \& Smith 1997).

NaSt1 has recently received renewed attention  principally
because it is extremely bright in the IR (K$=6.5$\,mag). Blum, 
DePoy \& Sellgren (1995), Tamblyn et al. (1996), Morris et al. (1996),
and Figer, McLean \& Najarro (1997) have
presented a spectral comparison  of various emission line objects, including
NaSt1, still adhering to its former WN10 or Ofpe/WN9 classifications. 
Indeed, NaSt1 is currently used as a late WN-type spectral standard,
particularly for IR studies of WR stars near the Galactic Centre,
despite its nature being uncertain.

In this paper we present new data for NaSt1 and consider its true nature.
Specifically, in 
Section~\ref{sect2} we report on new spectroscopy 
of NaSt1, obtained
at the Keck~I, William Herschel Telescope (WHT), and UK Infrared Telescope
(UKIRT). In Section~\ref{sect3} we discuss these new observations 
revealing a peculiar nebular appearance. In Section~\ref{sect4}
we use various techniques to determine the interstellar extinction 
and distance to NaSt1, while we obtain its nebular properties and 
abundances in Section~\ref{sect5}. We finally interpret our results
and discuss possible natures for NaSt1 in Section~\ref{sect6}.

\begin{figure*}
\epsfbox[31 354 516 541]{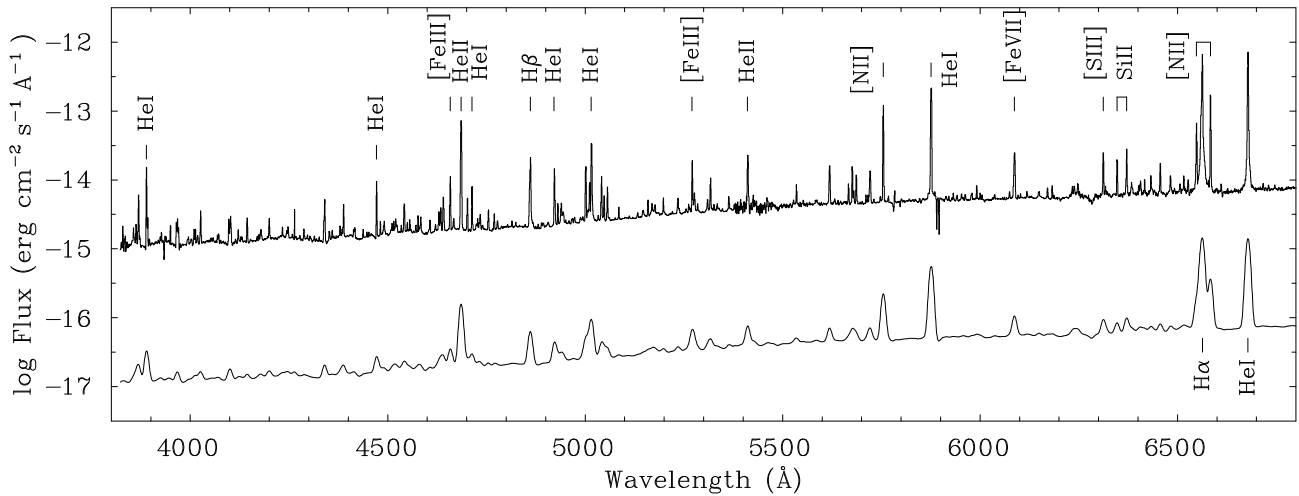}
\caption{WHT--ISIS flux-calibrated (erg\,cm$^{-2}$\,s$^{-1}$\,\AA)
spectrum of NaSt1 between $\lambda\lambda$3800--6800. Offset by --2 dex
is the identical spectrum, degraded to the spectral resolution ($\sim$6\AA) of 
the previous data set of Massey \& Conti (1983). Massey \& Conti assigned 
a WN10 spectral classification for NaSt1, on the basis of apparently broad 
stellar WR features, which are revealed here 
as exclusively narrow nebular features.}
\label{FIG1}
\end{figure*}

\begin{figure*}
\epsfysize=23.0cm \epsfbox[0 50 440 750]{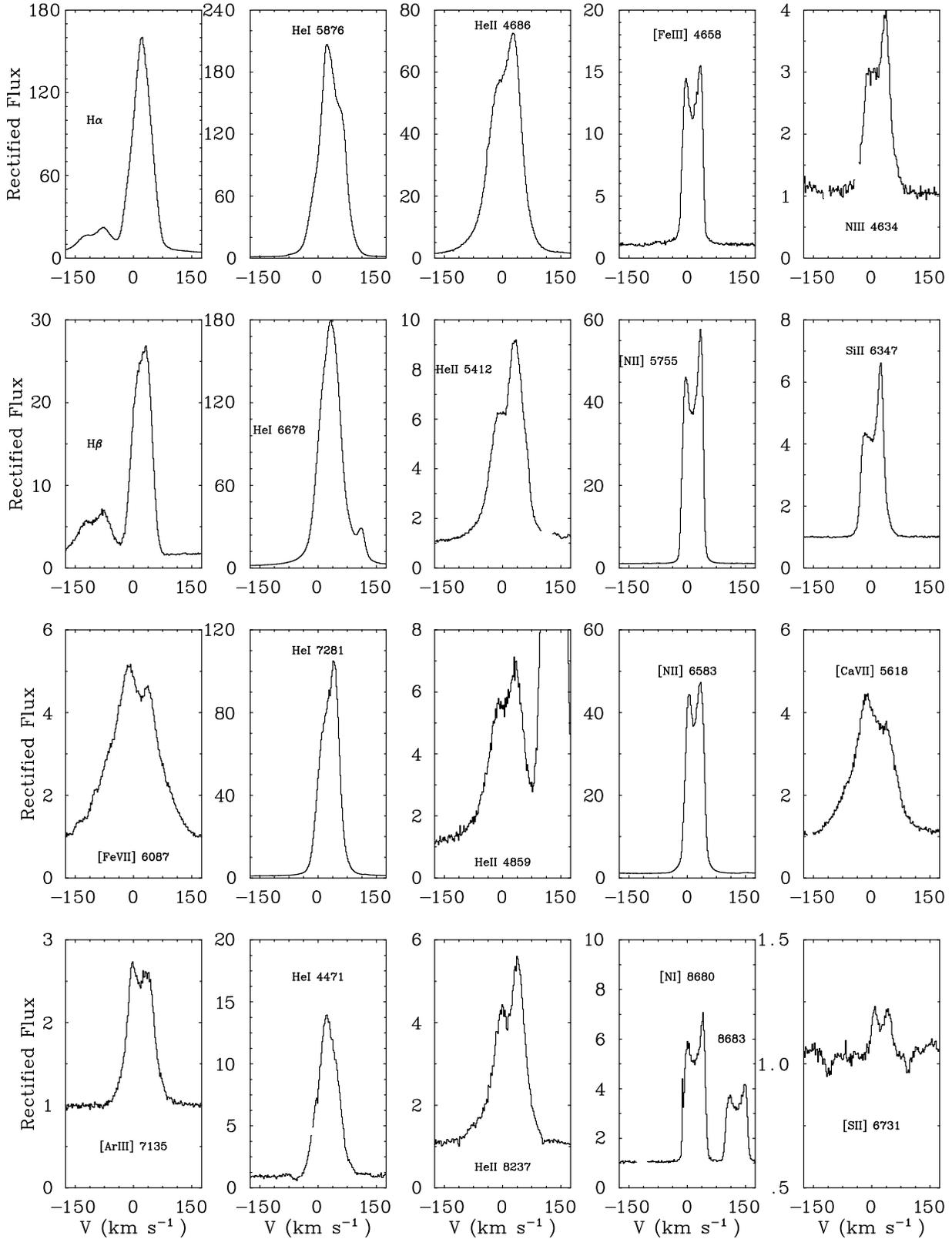}
\caption{Selected Keck I--HIRES nebular line profiles (radial velocities
are in LSR) for NaSt1, demonstrating the variety of morphologies seen.
Pickering He\,{\sc ii} profiles are present on the blue wings of 
the Balmer series (see e.g. He\,{\sc ii} $\lambda$4859).}
\label{FIG2}
\end{figure*}                

\section{Observations}\label{sect2}

We have obtained intermediate to high spectral resolution optical
and infrared spectroscopy of NaSt1 during 1994 July--October
at the 10m Keck~I, 4.2m WHT and 3.8m UKIRT telescopes.
These data were complemented by narrow-band filter imaging with the
auxiliary port of the WHT during August 1996.
The journal of our observations is presented in Table~\ref{TABLE1}.

\subsection{Optical observations}

Intermediate dispersion spectroscopy of NaSt1 between
$\lambda\lambda$3820--7030\,\AA\ were obtained at the 4.2m WHT,
during July--August 1994 using the 
the dual beam Intermediate dispersion Spectroscopic and Imaging
System (ISIS). These observations, obtained in good seeing (0.8$''$)
used 600~l/mm gratings on both arms of ISIS, with 
Tektronix and EEV CCDs (both 24$\mu$m pixels) on blue and red arms, 
respectively.
A 1$''$ slit width resulted in a spectral resolution of 1.6\AA\
(blue), and 1.7\AA\ (red) as determined from widths of 
CuAr and CuNe arc lines. 
The data were de-biased, divided by a normalised flat-field, and optimally
extracted using the {\sc pamela} (Horne 1986) routines within
{\sc figaro} (Shortridge et al. 1997). After wavelength calibration using
arcs obtained between stellar exposures, the spectra were
absolutely flux-calibrated using the standard star BD+28$^{\circ}$~4211.
Subsequent analysis was carried out within {\sc dipso} (Howarth et al. 1995).

Since the observed emission features were unresolved in our ISIS observations,
additional high spectral resolution observations of NaSt1 were kindly obtained 
for us by Dr M.H.~van~Kerkwijk  at the 10m Keck~I telescope, using 
the high resolution echelle spectrograph 
(HIRES) and a 2048$\times$2048 Tek CCD as the detector.
Observations at two wavelength 
settings during good conditions (0$''$.7 seeing) provided
near complete wavelength coverage between $\lambda\lambda$4350--8750 
at a 2.5\,pixel
spectral resolution of 0.08--0.15\AA, as measured from Th-Ar arc 
spectra. The CCD frames were bias-subtracted, and the echelle orders of 
NaSt1 and the 
flux and atmospheric standard Feige~110 were optimally extracted
using the software package {\sc echomop}  (Mills \& Webb 1994).  
Subsequent analysis was again performed using the {\sc figaro}
and {\sc dipso} packages.

Synthetic Johnson filter photometry performed on the 
WHT and Keck spectra give V=14.47 and 14.67 mag, respectively. 
Using 
the narrow-band photometric system derived for WR stars (Smith 1968) 
and convolving our flux calibrated observations with suitable 
Gaussian filters, we find $b$=16.91 mag, $v$=15.20 mag and $r$=13.99 mag 
for our WHT dataset. These compare reasonably well with previous
optical narrow band photometry ($b$=16.9 mag, $v$=15.4 mag) provided 
by Massey \& Conti (1983).

To complement our optical spectroscopy, imaging was carried out
with the WHT auxiliary port during August 1996 using the 1024$\times$1024
EEV CCD detector and several narrow-band filters,  covering He\,{\sc ii}
$\lambda$4686, He\,{\sc i} $\lambda$5876, H$\alpha$ + [N\,{\sc ii}] 
$\lambda$6583, and [N\,{\sc ii}] $\lambda$6583. The details
of the filters and exposure times employed are given in
Table~\ref{TABLE1}. The images were de-biassed and
flat-fielded prior to analysis. The average seeing measured from the
images was $0''.75$ with each CCD pixel corresponding to 0.22 arcsec.

\subsection{Infrared observations}

Our infrared NaSt1 observations were obtained at the 
3.8m UKIRT with the cooled grating spectrograph CGS4, the 300mm 
camera, the 75\,l/mm and echelle gratings and a 62$\times$58 InSb 
array in 1994 August covering selected regions in the 1.03--2.21$\mu$m range.
The observations were bias-corrected, flat-fielded, extracted and 
sky-subtracted using {\sc cgs4dr} (Daly \& Beard 1992). 
In order to remove atmospheric features, 
the observations were divided by
an appropriate standard star (whose spectral features were artificially
removed) observed at around the same time and similar air mass. 
Our echelle observations covered only He\,{\sc i} 2.058$\mu$m and 
were obtained at a spectral resolution of $\lambda$/$\Delta \lambda$=16\,000.

\section{Discussion of observations}\label{sect3}

In this section we discuss our WHT, Keck and UKIRT spectroscopy and imaging
of NaSt1. An extremely unusual nebular appearance 
is revealed, with no clear signature of stellar emission lines, 
arguing strongly against a WR identification.

\subsection{Optical spectroscopy}\label{optical}

We present our WHT-ISIS flux calibrated observations of NaSt1
in Fig.~\ref{FIG1}. The visual spectrum of NaSt1 shows 
a multitude of strong, narrow, low and high excitation nebular features 
superimposed on a clear continuum. From this figure it is clear that the 
nebular spectrum of NaSt1 is unusual. 
Very strong He\,{\sc i} features are observed relative to 
the Balmer series. 
In spite of the presence of very strong He\,{\sc ii} $\lambda$4686
emission -- indicating high temperatures or densities --
[O\,{\sc iii}] $\lambda$5007 is very weak and 
[N\,{\sc ii}] $\lambda6583$ is strong, suggestive of chemical 
peculiarities (see Sect.~\ref{sect6}).

In Fig.~\ref{FIG1} we include our WHT-ISIS observations
degraded to the resolution ($\sim$6\AA) of the sole
previously published optical spectroscopy of NaSt1 by 
Massey \& Conti (1983), obtained in 1982 September. 
On close inspection we find that the two datasets are essentially
identical, indicating that, whatever the true nature of NaSt1,
it has remained unchanged over the past decade.

{}From a spectral comparison with bona-fide cool
nitrogen sequence WR stars (specifically
the WN9 stars HDE\,313846 and BE381),
Massey \& Conti (1983) proposed a very late WN spectral 
classification for NaSt1 of WN10 on the basis that 
N\,{\sc ii} $\lambda\lambda$4654--67 emission was 
stronger than N\,{\sc iii} $\lambda\lambda$4634--41, and 
tentatively identified N\,{\sc i} $\lambda$5616,
in spite of very strong He\,{\sc ii} $\lambda$4686 emission.
(Smith et al. (1994) and Crowther \& Smith (1997) have recently 
updated the 
spectral classification of WN9--11 stars, with N\,{\sc i} absent).

The emission lines observed in NaSt1 are resolved in our
Keck~I HIRES data set, providing a potential means to unravelling 
its true nature. The stellar spectral 
features proposed by Massey \& Conti (1983) as N\,{\sc ii} 
$\lambda\lambda$4654--67 and N\,{\sc i} $\lambda$5616 which resulted in a 
late WN classification are revealed as nebular [Fe\,{\sc iii}] $\lambda$4658 
and [Ca\,{\sc vii}] $\lambda$5619, respectively. Indeed, {\it all optical 
spectral features appear to be of nebular, rather than stellar, origin}.
Permitted and forbidden lines cover a wide range  in excitation, and include
H\,{\sc i}, He\,{\sc i-ii},  N\,{\sc i-iii}, [N\,{\sc ii}], [Ne\,{\sc
iii-iv}], Mg\,{\sc i-ii}, Si\,{\sc ii},
[S\,{\sc ii-iii}], [Ar\,{\sc iii-v}],  [Ca\,{\sc v-vii}], [Fe\,{\sc
ii-vii}], [Ni\,{\sc ii-iii}]). 

We present selected line profiles from the Keck dataset in
Fig.~\ref{FIG2}, covering a wide range of morphologies, which we 
now discuss. A full line list is provided in Table~\ref{TABLE2}.
\begin{enumerate}
\item {\bf Balmer series}: H$\alpha$ and H$\beta$ show very similar
emission line profiles with a double component structure, comprising
a major component with FWHM$\sim$50\,km\,s$^{-1}$ at line centre
plus a minor component centred at $\sim$$-$115\,km\,s$^{-1}$,
which we attribute to He\,{\sc ii} since these features
have the correct velocity displacement, the appropriate strength
and an identical structure to adjacent Pickering series members.
\item {\bf He\,{\sc i}}: The numerous He\,{\sc i} emission lines 
in our observations show a variety of morphologies. Most He\,{\sc i}
profiles, such as $\lambda$7281 in Fig.~\ref{FIG2} show double Gaussian, 
asymmetric profiles (FWHM$\sim$33\,km\,s$^{-1}$), separated by 
$\sim$32\,km\,s$^{-1}$, with stronger emission in the red component. 
Other He\,{\sc i} lines such as $\lambda$4713, $\lambda$5016 show a 
broader profile that can be deconvolved into two components, again 
stronger on the red side (FWHM$\sim$43\,km\,s$^{-1}$ separated 
by $\sim$27\,km\,s$^{-1}$). Exceptions include $\lambda$5876
(see Fig.~\ref{FIG2}) which 
shows a double Gaussian profile with greater blue emission
and $\lambda$6678 which has a single 
symmetric profile with FWHM=62\,km\,s$^{-1}$. 
\item {\bf He\,{\sc ii}}: The profiles 
(e.g. $\lambda4686$, $\lambda4859$, $\lambda5412$, $\lambda$8237
in Fig.~\ref{FIG2}) are asymmetric, and can readily be deconvolved into
two components with FWHM$\sim$50\,km\,s$^{-1}$, separated by
$\sim$50\,km\,s$^{-1}$ with blue to red strengths of 2:3.
\item {\bf Low excitation forbidden lines}: Spectral lines from these
transitions are fairly common 
and include [N\,{\sc i}] $\lambda$8680, 
[N\,{\sc ii}] $\lambda$5755, $\lambda$6583, [S\,{\sc ii}] $\lambda$6731,
[Fe\,{\sc iii}] $\lambda$4658 presented in Fig.~\ref{FIG2}.
These profiles can be reproduced with double Gaussian fits of
similar strength, with intrinsic FWHM$\sim$24\,km\,s$^{-1}$, and separated
by 28--31\,km\,s$^{-1}$.
\item {\bf High excitation forbidden lines}: Spectral features due to
[Ca\,{\sc v-vii}], [Fe\,{\sc v-vii}] are observed 
(see Fig.~\ref{FIG2}). While the broad shape of these lines
can be reproduced by a single Gaussian fit 
(FWHM$\sim$140\,km\,s$^{-1}$),  a double peaked structure is observed, 
with each  component separated by $\sim$50\,km\,s$^{-1}$.
\item {\bf Permitted metal lines}: Examples 
include N\,{\sc iii} $\lambda\lambda$4634--41 and Si\,{\sc ii}
$\lambda$6347--71, shown in Fig.~\ref{FIG2}. These are of 
similar shape to the He\,{\sc ii} profiles, with asymmetric, double
peaked profiles (FWHM$\sim$28--36\,km\,s$^{-1}$), 
separated by 36--41\,km\,s$^{-1}$,
showing greater emission in the red component.
\end{enumerate}

\begin{figure}
\epsfxsize=8.0cm \epsfbox[70 200 415 565]{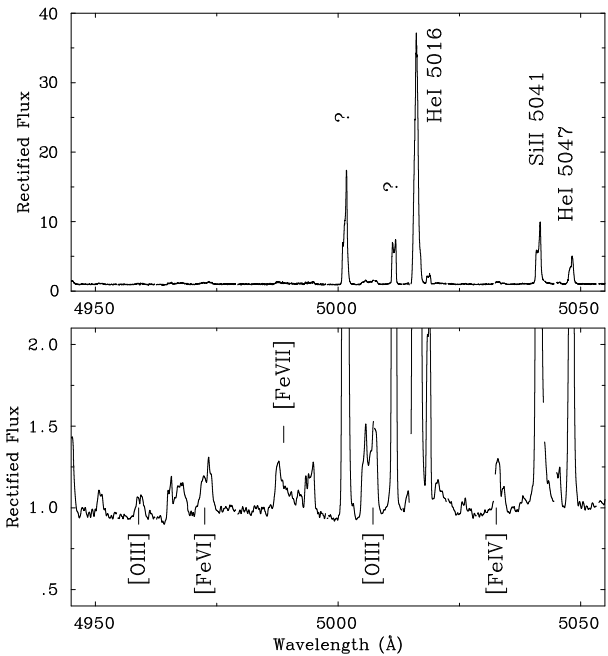}
\caption{A portion of our Keck I--HIRES spectrum demonstrating 
the weakness of the [O\,{\sc iii}] $\lambda\lambda$4959, 5007 lines 
(compare with He\,{\sc i} $\lambda$5016).}
\label{FIG3}
\end{figure}

In Fig.~\ref{FIG3} we present a small portion of our HIRES spectrum in
the region of the [O\,{\sc iii}] $\lambda$5007, $\lambda$4959
lines. These lines, usually amongst the strongest nebular lines in hot
Planetary Nebulae (PNe), are extremely weak in NaSt1.  Indeed, our line
measurements indicate that the feature at $\lambda$5007 is blended,
since it is seven times stronger than $\lambda$4959 (the
theoretical line ratio is 2.9).  Although our spectroscopic
observations do not extend to the [O\,{\sc ii}] doublet at
$\lambda3727$, we have been provided with intermediate dispersion
observations of NaSt1 from L.F.~Smith extending to $\lambda$3300. From
this data set, negligible emission is observed at [O\,{\sc ii}]
$\lambda$3727. The high reddening towards NaSt1 (Sect. 4.1), however,
means that no useful limit can be determined for the strength of this
feature. Instead, we have searched for the red [O\,{\sc ii}] lines at
7320 and 7330\,\AA\ in the Keck dataset. The latter occurs in the
inter-order gap but the former is detected and has approximately
the same strength as the blended [O\,{\sc iii}] $\lambda5007$ line.  
The weakness of
the oxygen lines therefore indicates that the oxygen content of NaSt1
is very low. Likewise, we can make inferences about the carbon content
since, by comparison with planetary nebulae and the
presence of strong He\,{\sc ii} $\lambda4686$ in NaSt1, we would
expect to see the C\,{\sc iv} recombination lines at $\lambda4660$, 
$\lambda\lambda$5801--12,
but these are absent in our spectra. The C\,{\sc ii}
line at $\lambda4267$ is also absent, suggesting that NaSt1 is
carbon deficient.

\begin{figure}
\epsfxsize=7.0cm \epsfbox[72 230 470 590]{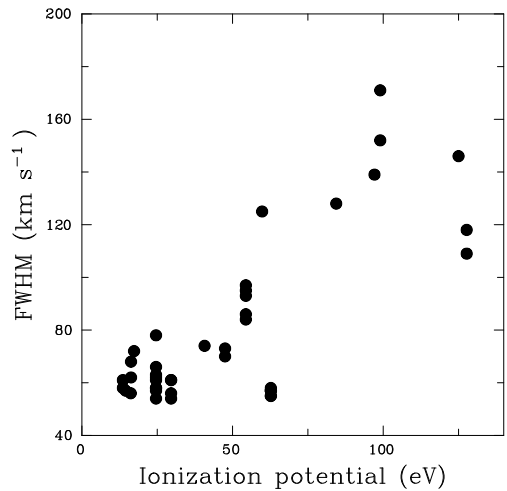}
\caption{Correlation between FWHM (km\,s$^{-1}$) and
ionization potential (eV) for nebular lines observed in Keck-HIRES
observations of NaSt1.}
\label{FIG4}
\end{figure}

In Fig.~\ref{FIG4} we present line FWHM (km\,s$^{-1}$) 
versus ionization potential (in eV) for representative ions in 
our HIRES spectra 
of NaSt1, revealing a broad correlation, suggesting lines of different
excitations are formed within different regions of the nebula, as seen, for
example, within symbiotic novae (e.g. V1016 Cyg, Schmid \& Schild 1990). 

\begin{figure}
\epsfxsize=8.0cm \epsfbox[110 125 440 700]{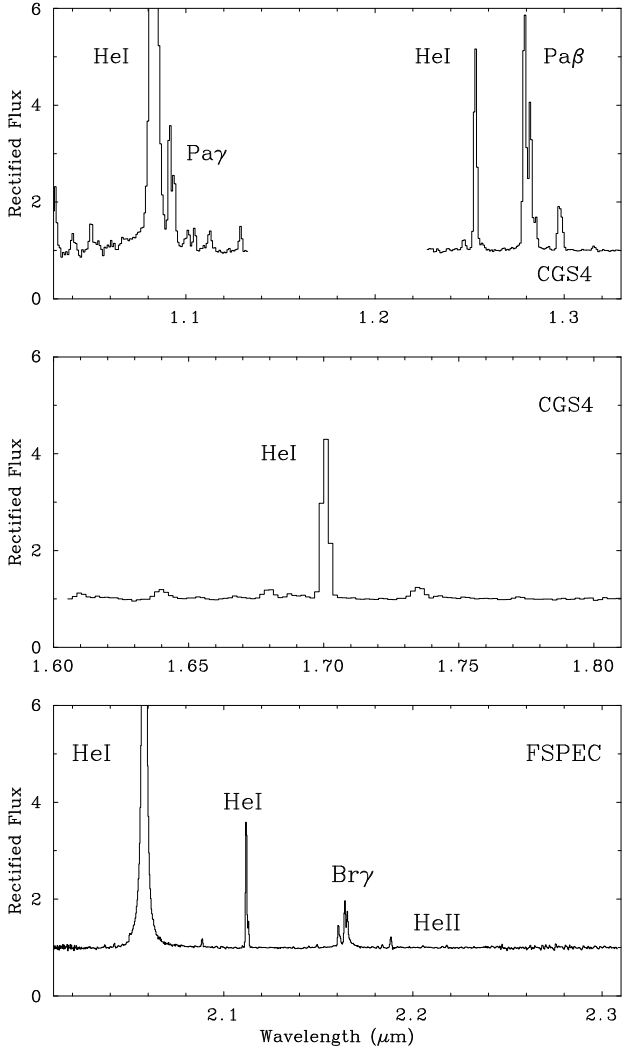}
\caption{Infrared UKIRT/CGS4 spectroscopy of NaSt1,
plus the Steward Bok/FSPEC 2.0-2.3$\mu$m spectra of Tamblyn et al. (1996).
In our low resolution I-band spectrum the He\,{\sc i} 1.083$\mu$m
profile reaches 300 (!) times the local continuum.}
\label{FIG5}
\end{figure}

\subsection{Infrared spectroscopy}\label{ir_spect}

The low resolution 1.0--1.8$\mu$m UKIRT/CGS4 spectrum of NaSt1 is shown in
Figure~\ref{FIG5} together with 
a higher resolution 1.99--2.31$\mu$m
spectrum obtained at the Steward Bok 2.3m/FSPEC by Tamblyn et al. (1996).
Once again, numerous nebular emission lines, principally attributable
to H\,{\sc i} and He\,{\sc i-ii} are observed, with weak features
tentatively identified as nitrogen and iron (Table~\ref{TABLE3}).
While our observations are generally of insufficient quality
to resolve individual features, the hydrogen and helium components at 
2.165$\mu$m are resolved (He\,{\sc i} 2.162$\mu$m + He\,{\sc ii} 
2.1646$\mu$m + Br$\gamma$) in the FSPEC data set, as are the 
Balmer-Pickering series in the optical. 

\begin{figure}
\epsfxsize=8.0cm \epsfbox[110 320 440 500]{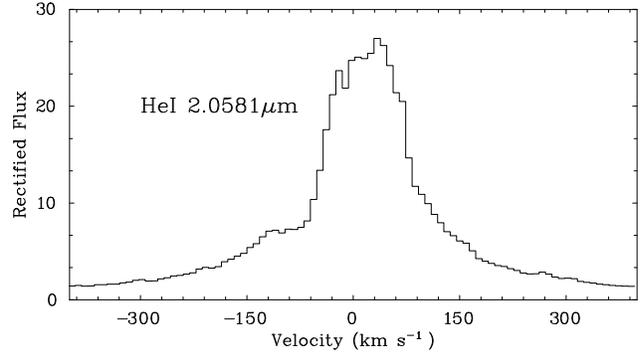}
\caption{High resolution UKIRT/CGS4 echelle spectrum of NaSt1,
at He\,{\sc i} 2.0581$\mu$m, demonstrating the extended emission
wings in this line, which may be the sole wind velocity indicator 
of  the ionizing source.}
\label{FIG6}
\end{figure}

In Fig.~\ref{FIG6} we present the high resolution 
He\,{\sc i} 2.058$\mu$m UKIRT echelle profile. These observations 
reveal a strong, emission feature with wings extending to 
$\sim$300\,km\,s$^{-1}$.  This feature can be reproduced with 
a double Gaussian fit comprising a narrow central feature with 
FWHM$\sim$110\,km\,s$^{-1}$ plus a second, broad component of 
similar flux with FWHM$\sim$360\,km\,s$^{-1}$. This profile 
represents  the only potential feature that  has a stellar 
origin. Consequently, $\sim$300\,km\,s$^{-1}$ may relate to 
the underlying outflow wind velocity. 

\subsection{Optical imaging}\label{opt_image}

We now discuss our narrow-band imaging of NaSt1 obtained at the
auxiliary port of the WHT. To date, the only published image of NaSt1
is an H$\alpha$+[N\,{\sc ii}] image from Miller \& Chu (1993).  
This image shows no nebular emission associated with the star 
although the pixel size of $0''.98$ means that a nebula close 
to the central star would have been missed.
Williams et al. (1987) find evidence from the IR flux distribution for
a dusty circumstellar shell associated with NaSt1.

\begin{figure*}
\epsfbox[54 304 478 723]{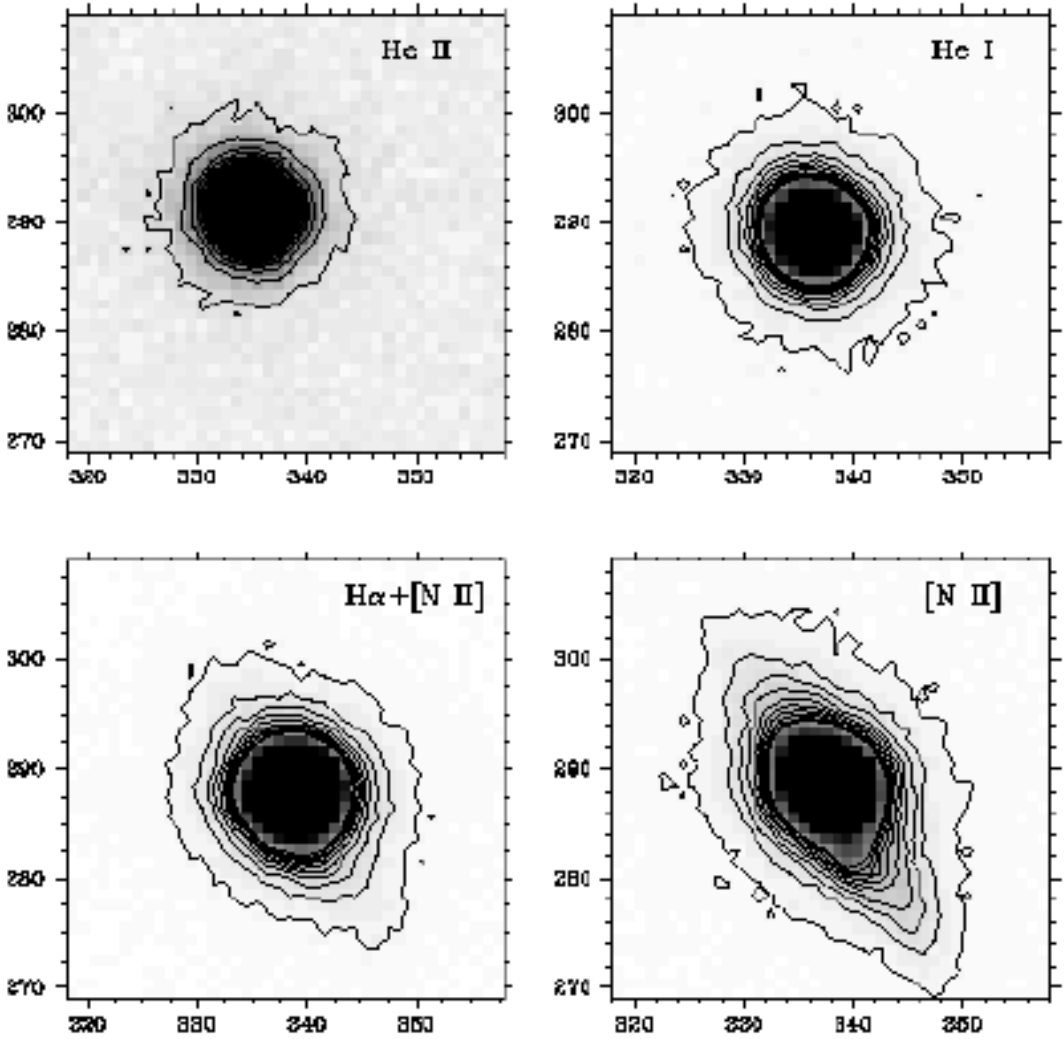}
\caption{Narrow-band images of NaSt1 obtained at the WHT. Each image is $40
\times 40$ pixels ($=8''.8 \times 8''.8$) where 1 CCD pixel is 0.22 arcsec.
The contours are plotted at intervals of 100 counts with the lowest
contour at 70 counts.}
\label{FIG7}
\end{figure*}

In Fig.~\ref{FIG7} we show the four narrow-band images of NaSt1 with
contours superimposed. It is immediately obvious that a nebula is
detected in the [N\,{\sc ii}] $\lambda$6583 image.  It is
elliptical in shape with the major axis at a position angle of
$30^\circ$. The lengths of the major and minor axes (using the lowest
contours shown in Fig.~\ref{FIG7}) are 8.5 and 5.1 arcsec
respectively. The nebula is also seen in the 
H$\alpha+$[N\,{\sc ii}] $\lambda$6565 image. In contrast, there 
is no hint of any extension in the He\,{\sc ii} and He\,{\sc i} 
images at the measured seeing of $0''.75$.

The [N\,{\sc ii}] $\lambda6583$ line is also spatially extended
in the Keck HIRES spectra which were obtained at a position angle of
$0^\circ$. Examination of the [N\,{\sc ii}] profile shows that the
relative strengths of the blue and red components vary as a function
of spatial position. At the position of the continuum, they 
have equal strengths and are separated by
30~km\,s$^{-1}$; to the north, the blue component dominates; and to
the south, the reverse occurs with the red component much stronger.
At all positions, both components are always seen and their velocities
are constant with no sign of the two components merging at the edge
of the nebula. The dynamics are therefore inconsistent
with a simple expanding shell, but suggest a more complex geometry.

Pre-empting the derived distance for NaSt1 in the next
section, an average diameter for the [N\,{\sc ii}] emitting region
of 6.8 arcsec
corresponds to a physical size of 0.033 pc (or 6,800 AU) at a 
distance of 1\,kpc, or 0.11 pc (22,400 AU) at a distance of 3.3\,kpc. 
If we assume that the characteristic expansion velocity associated
with the outflowing material is 15 km\,s$^{-1}$, we derive a dynamical 
timescale for the [N\,{\sc ii}] emitting region of 1,100--3,600\,yr.

\section{Extinction and distance towards NaSt1}\label{sect4}

We now use our spectroscopic observations presented above
to investigate the interstellar extinction and distance 
to NaSt1.

\subsection{Energy distribution and interstellar extinction}

\begin{figure}
\epsfxsize=8.8cm \epsfbox[59 263 425 521]{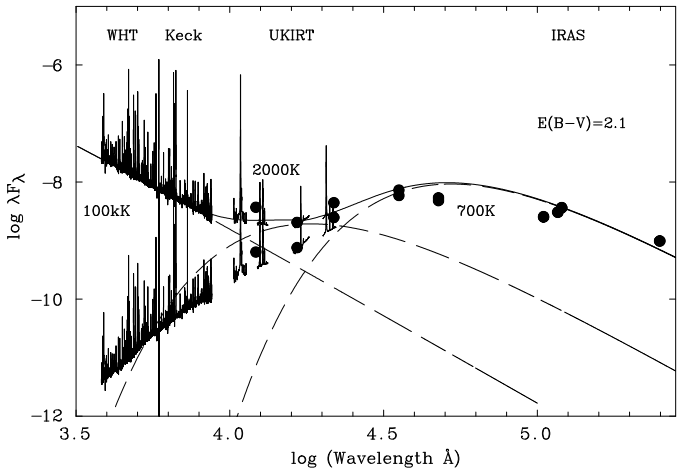}
\caption{Observed and de-reddened WHT, Keck and UKIRT spectrophotometry
plus IR photometry (Williams et al. 1987; Shylaja \& Anandarao 1993) for 
NaSt1 (log [$\lambda$$F_{\lambda}$] vs. log $\lambda$). NaSt1 is the
IRAS source IRAS18497+0056. Individual black bodies 
are shown as dashed-lines (100kK, 2000K, 700K) with a solid line indicating
their combined energy distribution.}
\label{FIG8}
\end{figure}

The observed 0.4--15$\mu$m flux distribution
for NaSt1 is presented in Fig.~\ref{FIG8}. The energy 
distribution is extremely red, and 
peaks around the L-band, suggestive of a very high interstellar
reddening. We defer a comparison between the observed
IR colours for NaSt1 with other emission line objects until
Sect.~\ref{comparison}.

We use two methods to determine the interstellar extinction
to NaSt1; applying Case~B recombination theory 
(Storey \& Hummer 1995) to the  observed 
H\,{\sc i} line strengths, and the strength of
observed Diffuse Interstellar Band (DIB) absorption lines.

For the first method, we have assumed an electron temperature
$T_{e}$=10,000K and electron density $N_{e}$=10$^{4}$\,cm$^{-3}$ and
used the H$\alpha$, P13 and P16 fluxes relative to H$\beta$ from our
Keck/HIRES observations. Higher Balmer series were not used since
measurements of these features are restricted to less reliable
WHT/ISIS observations.  We obtain $c$(H$\beta$)=3.06$\pm0.08$,
implying E$_{\rm B-V}$=2.1$\pm$0.1 mag.  In Sect.~\ref{sect5}, we derive
higher values of $T_{e}$ and $N_{e}$ but these have a negligible effect
on the derived value of E$_{\rm B-V}$.

NaSt1 lies along the line-of-sight to the Aquila Rift (200$\pm$100\,pc) 
which Dame \& Thaddeus (1985) suggest has a low mean visual extinction
of about 2 magnitudes.  It appears that the majority of the
extinction we observe arises from diffuse material lying behind this cloud, 
and therefore that the standard mean value of 
$R$ (=$A_{\rm V}$/E$_{\rm B-V}$)=3.1 may be reasonable 
(i.e. $A_{\rm V}$=6.5$\pm$0.3 mag).

Diffuse Interstellar Band (DIB) features are readily visible
in our Keck/HIRES spectra of NaSt1 and allow an independent 
$E_{\rm B-V}$ determination (see also Le~Bertre \& Lequeux 1993). 
In particular, equivalent widths of 
certain DIBs ($\lambda$5797, $\lambda$5849)
are known to scale fairly linearly with $E_{\rm B-V}$ (Herbig 1995). 
We have measured equivalent widths for these DIBs and compared them
with those quoted by Herbig (1995) along the (standard) line-of-sight 
to HD\,184143. We obtain equivalent widths of 377 and 135 m\AA\ for 
$\lambda$5797 and $\lambda$5849, implying $E_{\rm B-V}$=2.00 and 
2.04\,mag,
in excellent
agreement with our value derived from Case~B recombination theory.

The de-reddened flux distribution for NaSt1 is shown in
Fig.~\ref{FIG8}. Clearly, the intrinsic optical flux
distribution from NaSt1 is very blue, indicating a very hot
ionizing source for the nebula, as illustrated by the 100,000K blackbody 
flux  distribution in the figure. Our optical spectroscopy does
not allow us to distinguish between temperatures for the ionizing 
source in the range 30kK (if $E_{\rm B-V}$=2.0 mag) to $\ge$200kK (if
E$_{\rm B-V}$=2.2 mag). The mid-IR energy distribution 
can be approximated with a warm blackbody of 700K 
(van der Hucht et al. 1984;  Williams et al. 1987), while a further
blackbody of $\sim$2000K is required to reproduce the intrinsic
near-IR flux distribution, The combined effect of 
our three blackbodies is indicated as a solid line in Fig.~\ref{FIG8}.

\subsection{Distance}

We use several techniques to estimate the distance, based on the LSR
radial velocity of NaSt1 and interstellar material, and from
previously obtained distance-reddening relations along this
line-of-sight.

\begin{figure}
\epsfxsize=8.5cm \epsfbox[65 125 530 666]{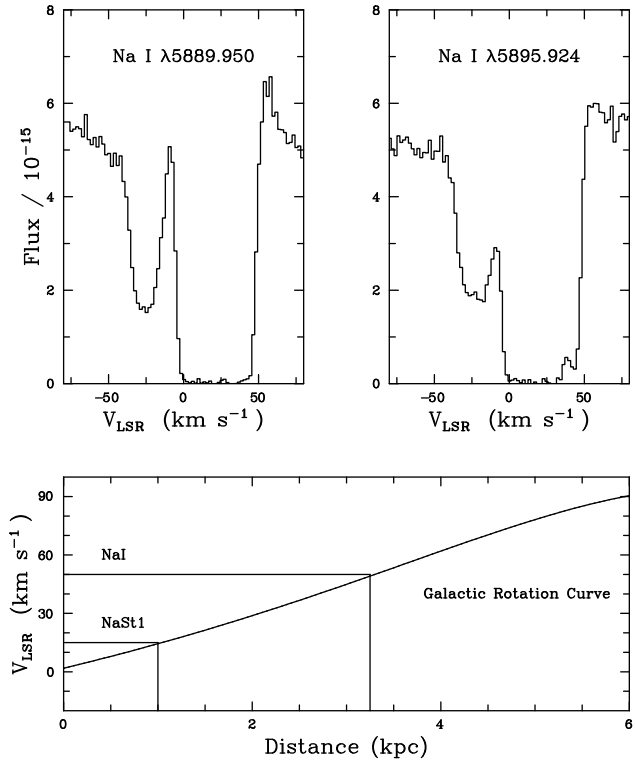}
\caption{The HIRES spectrum of NaSt1 showing the interstellar 
Na\,{\sc i} D lines and the LSR radial velocity predicted 
as a function of distance for the line-of-sight towards NaSt1 
using the Galactic rotation curve of Brand \& Blitz (1993).
The systemic velocity of the emission lines and the velocity structure
of the Na\,{\sc i} $\lambda\lambda$5890,5896 absorption
lines indicate a distance of 1--3.3\,kpc.}
\label{FIG9}
\end{figure}

To estimate the systemic LSR radial velocity for the emission lines
observed in NaSt1, we have measured the velocities of thirteen
emission lines which show red and blue components (see
Fig.~\ref{FIG2}) in the Keck HIRES dataset. We derive $V_{\rm LSR}=-4\pm6$
and $+34\pm5$~km~s$^{-1}$ for the blue and red components, giving a
systemic $V_{\rm LSR}$ of $+15$~km~s$^{-1}$ for NaSt1. The interstellar
Na\,{\sc i} lines are shown in Fig.~\ref{FIG9} and have a strong,
broad component centred on $V_{\rm LSR}=+25$~km~s$^{-1}$, and extending to
$+50$~km~s$^{-1}$, with a weaker component at $-25$~km~s$^{-1}$.

The LSR radial velocity as a function of distance for the
line-of-sight towards NaSt1 is also shown in Fig.~\ref{FIG9}, using
the Galactic rotation curve of Brand \& Blitz (1993) and a
Galactocentric distance of 8.5 kpc for the Sun. The velocity of
$+15$~km~s$^{-1}$ derived from the emission lines indicates a distance
of $\approx$1\,kpc for NaSt1.  In contrast, the main Na\,{\sc i}
absorption feature, extending up to $+50$~km~s$^{-1}$ suggests a
distance of $\sim$3.3\,kpc. The first estimate assumes that NaSt1 is
participating in the Galactic rotation and has no peculiar velocity.
The second estimate assumes that the Na\,{\sc i} absorption originates
from diffuse interstellar clouds along the line-of-sight which are
co-rotating with the Galaxy. It is possible that some of this absorption
arises in the circumstellar shell associated with NaSt1 since the red
emission component is at $V_{\rm LSR}=+34$~km~s$^{-1}$. On the other hand,
there is no blue-shifted emission with velocities as negative as
the additional Na\,{\sc i} component at $-$25~km~s$^{-1}$.

Alternatively, Cappellaro et al. (1994) have presented an approximate
distance-reddening relation for objects along a narrow
line-of-sight common to NaSt1
based on normal stars  and PNe. Using their approximate relation 
our $E_{\rm B-V}$ implies a significantly larger distance of 7$\pm$1\,kpc. 
It is possible, however, that the high extinction towards NaSt1 results
from dusty circumstellar material rather than interstellar
material. 

In summary, the high reddening of $E_{\rm B-V}=2.1$\,mag derived from
H\,{\sc i} line ratios indicates a large distance of $\sim 7$\,kpc but
some of this reddening could be local to NaSt1.
Conversely, the radial
velocities of the emission and interstellar absorption lines indicate
smaller distances of 1--3.3\,kpc and by inference, that most of the
reddening is circumstellar.  Dame \& Thaddeus (1985) list nine O stars
lying behind the Aquila Rift with distances in the range 1--6\,kpc and
extinctions of $A_{\rm V}$=2.4--4.2 mag, considerably less than the
value derived for NaSt1. We have a slight preference for the distances
based on kinematic arguments and will assume a distance to NaSt1 of
1--3.3\,kpc.

This distance range
yields an absolute visual magnitude for NaSt1 of $M_{\rm V}$=$-$1.9 to
$-$4.5\,mag. Assuming a bolometric  correction in the 
range $-$3 to $-$7\,mag, appropriate for stars with temperatures of
30,000K to 200,000K, the intrinsic luminosity of the hot 
component lies in the range log\,($L/L_{\odot}$)=3.9 to 6.5. 
We can also obtain a luminosity from the IRAS mid-IR
photometry, which reflects warm re-radiated dust. For a distance
of 1 to 3.3\,kpc we obtain log\,($L/L_{\odot}$)=2.6 to 3.6 (assuming 
a 700K blackbody normalized at the observed 12$\mu$m flux). Clearly,
although extremely bright at IR wavelengths, the entire 
bolometric luminosity of NaSt1 is not re-radiated in the IR.

\begin{table}
\centering
\begin{minipage}{80mm}
\caption{Observed ($F_{\lambda}$) and de-reddened ($I_{\lambda}$)
emission lines fluxes for NaSt1 from Keck~I--HIRES observations,
assuming $E_{\rm B-V}$ = 2.1 mag ($c$(H$\beta$) = 3.06).
As is usual, fluxes shown are relative to H$\beta$ (=100) for which 
$F_{H \beta}$=4.847$\times 10^{-14}$ erg\,cm$^{-2}$\,s$^{-1}$ and
$I_{H \beta}$=5.433$\times 10^{-11}$ erg\,cm$^{-2}$\,s$^{-1}$.
In spectral regions for which Keck~I--HIRES data are absent, we 
utilise WHT--ISIS values (labelled WHT), while regions without
spectroscopy, lying between HIRES echelle orders, are indicated 
with `$\ldots$'. Lines which are heavily blended (bl.) or broad 
(br.) are indicated.}
\label{TABLE2}
\begin{tabular}{@{}clrcrrl}
$\lambda_{\rm obs}$&\multicolumn{3}{c}{Identification} &$F_{\lambda}$ &$I_{\lambda}$ &Note\\
(\AA)              & Feature && $\lambda_{\rm air}$     &H$\beta$=100& H$\beta$=100&\\
\hline
3856.2   & Si\,{\sc ii} &1 & 3856.0   &   1.9 &  9.8&WHT\\
3862.8   & Si\,{\sc ii} &1 & 3862.6   &   2.9 & 14.6&WHT\\
3869.1   & Ne\,{\sc iii}&3F& 3868.8   &  20.9 &105.9&WHT\\
3888.9   & He\,{\sc i}  &  & 3888.6   &  51.0 &251.5&WHT\\
3967.6   & H\,{\sc i}   &  & 3968.5   &  15.4 & 68.8&WHT\\
         & He\,{\sc i}  &5 & 3964.7   &       &     &WHT\\
4009.3   & He\,{\sc i}  &55& 4009.3   &   2.9 & 12.0&WHT\\
4026.6   & He\,{\sc i}  &19& 4026.6   &   9.9 & 40.0&WHT\\
4068.9   & S\,{\sc ii}  &1F& 4068.9   &   1.3 &  4.9&WHT\\
4076.7   & S\,{\sc ii}  &1F& 4076.2   &   0.4 &  1.5&WHT\\
4097.0   & N\,{\sc iii} & 1& 4097.3   &   7.1 & 25.7&WHT\\
4101.7   & H$\delta$    &  & 4101.7   &  10.3 & 37.0&WHT\\
         & N\,{\sc iii} & 1& 4103.3   &       &     &WHT\\
4121.4   & He\,{\sc i}  &16& 4120.8   &   3.0 & 10.5&WHT\\
4144.4   & He\,{\sc i}  &53& 4143.8   &   6.5 & 21.9&WHT\\
4178.9   & Fe\,{\sc ii} &28& 4178.9   &   2.7 &  8.6&WHT,bl?\\
4199.7   & He\,{\sc ii} &  & 4199.9   &   7.5 & 23.1&WHT\\
4233.3   & Fe\,{\sc ii} &27& 4233.2   &   2.4 &  7.0&WHT\\
4247.7   &              &  &          &   3.0 &  8.5&WHT\\
4287.7   & Fe\,{\sc ii} &7F& 4287.4   &   2.6 &  6.9&WHT\\
4339.9   & H$\gamma$    &  & 4340.8   &  21.4 & 52.3&WHT\\
         & He\,{\sc ii} &  & 4338.7   &       &&\\
         & O\,{\sc iii} &3F& 4363.2   &$<$0.5 &$<$1.2&\\
4388.0   & He\,{\sc i}  &51& 4388.1   &  14.1 & 31.8&\\
4413.8   & Fe\,{\sc ii} &7F& 4413.8   &   2.0 &  4.3&\\
4416.3   & Fe\,{\sc ii} &6F& 4416.3   &   2.1 &  4.5&\\
4437.6	 & He\,{\sc i}  &50& 4437.5   &   2.8 &  5.8&\\
4446.9   & N\,{\sc ii}  &  & 4447.0   &   1.2 &  2.4&\\
4471.7   & He\,{\sc i}  &14& 4471.5   &  39.7 & 78.7&\\
4481.2	 & Mg\,{\sc ii}?&4 & 4481.3   &   4.7 &  9.1&\\
4491.4   & Fe\,{\sc ii} &37& 4491.4   &   2.0 &  3.8&\\
4511.0   & Fe\,{\sc ii} &6F& 4509.6   &   3.4 &  6.2&\\
4515.3   & Fe\,{\sc ii} &37& 4515.3   &   2.5 &  4.5&\\
4520.1   & Fe\,{\sc ii} &37& 4520.2   &   3.3 &  5.9&\\
4533.6   & Fe\,{\sc ii} &6F& 4533.0   &   2.2 &  3.9&\\
4541.6   & He\,{\sc ii} &  & 4541.7   &  14.8 & 25.7&\\
4549.5   & Fe\,{\sc ii} &38& 4549.5   &   2.9 &  5.0&\\
4554.8   &              &  &          &   3.0 &  5.1&\\
4571.0   & Mg\,{\sc i}  &1 & 4571.1   &   2.2 &  3.6&\\
4576.4   & Fe\,{\sc ii} &38& 4576.3   &   1.5 &  2.5&\\
4583.8   & Fe\,{\sc ii} &38& 4583.8   &   5.0 &  8.1&bl.\\
4607.0   & Fe\,{\sc iii}&3F& 4607.0   &   5.0 &  7.8&\\
4620.6   & Fe\,{\sc ii} &38& 4620.5   &   1.5 &  2.2&\\
4629.3   & Fe\,{\sc ii} &37& 4629.1   &   5.7 &  5.1&\\
4630.6   &              &  &          &   1.8 &  2.7&\\
4634.2	 & N\,{\sc iii} &2 & 4634.1   &  10.2 & 15.1&\\
4640.7	 & N\,{\sc iii} &2 & 4640.6   &  20.2 & 29.6&\\
4658.1   & Fe\,{\sc iii}&3F& 4657.7   &  42.5 & 60.5&\\
4666.9   & Fe\,{\sc iii}&3F& 4667.0   &   5.7 &  8.0&\\
4685.8   & He\,{\sc ii} &  & 4685.8   & 403.3 &546.5&\\
4701.6   & Fe\,{\sc iii}&3F& 4701.6   &  18.4 & 24.2&\\
4713.4   & He\,{\sc i}  &  & 4713.2   &  36.9 & 47.7&\\
4724.1   & Ne\,{\sc iv} &1F&4724.1    &   3.7 &  4.7&br.\\
4731.4   & Fe\,{\sc ii} &43&4731.4    &   2.2 &  2.7&\\
\hline
\end{tabular}
\end{minipage}
\end{table}

\addtocounter{table}{-1}

\begin{table}
\centering
\begin{minipage}{80mm}
\caption{(continued)}
\begin{tabular}{@{}clrcrrl}
$\lambda_{\rm obs}$&\multicolumn{3}{c}{Identification} &$F_{\lambda}$ &$I_{\lambda}$ &Note\\
(\AA)              & Feature && $\lambda_{\rm air}$     &H$\beta$=100& H$\beta$=100&\\
\hline
4733.9   & Fe\,{\sc iii}&3F& 4733.9   &   4.6 &  5.7&\\
4739.9   &Ar\,{\sc iv}  &1F&4740.1    &   1.9 &  2.3&br\\
4754.8   & Fe\,{\sc iii}&3F& 4754.8   &   7.8 &  9.4&\\
4769.5   & Fe\,{\sc iii}&3F& 4769.6   &   6.1 &  7.2&\\
4777.7   & Fe\,{\sc iii}&3F& 4777.9   &   2.8 &  3.2&\\
4814.5   & Fe\,{\sc ii}&20F& 4814.5   &   1.9 &  2.1&\\
4824.1   &              &  &          &   1.2 &  1.3&\\
4859.4   & He\,{\sc ii} &  & 4859.4   &  34.9 & 34.9&\\
4861.4   & H$\beta$     &  & 4861.2   & 100.0 &100.0&\\
4864.1   &              &  &          &   8.6 &  8.6&bl.\\
4889.6   & Fe\,{\sc ii} &4F& 4889.6   &   2.2 &  2.1&\\
4921.9   & He\,{\sc i}  &48& 4921.9   &  66.4 & 59.9&\\
4923.8   & Fe\,{\sc ii} &42& 4923.9   &   4.2 &  3.8&\\
4930.6   & Fe\,{\sc iii}&1F& 4931.0   &  10.5 &  9.3&\\
4939.1   & Ca\,{\sc vii}&1F& 4940.3   &  13.8 & 12.0&\\
4944.4   & Fe\,{\sc vii}&2F& 4942.3   &  10.3 &  9.0&bl.\\
4958.9   & O\,{\sc iii} &1F& 4959.3   &   1.2 &  1.0&\\
4967.1   & Fe\,{\sc vi} &2F& 4967.1   &   2.9 &  2.4&br.\\
4972.4   & Fe\,{\sc vi} &2F& 4972.5   &   3.9 &  3.2&br.\\
4987.3   & Fe\,{\sc vii}&2F& 4988.8   &   3.9 &  3.2&\\
4993.9   &              &  &          &   2.8 &  2.2&\\
5001.1   &              &  &          &  65.0 & 51.1&\\
5007.2   & O\,{\sc iii} &1F& 5006.9   &$<$8.4 &$<$6.5&bl.\\
5011.3   &              &  &          &  29.4 & 22.7&\\
5015.8   & He\,{\sc i}  &4 & 5015.7   & 182.7 &140.1&\\
5018.4   & Fe\,{\sc ii} &42& 5018.4   &   6.6 &  5.0&\\
5032.6   & Fe\,{\sc iv} &-F& 5032.3   &   2.3 &  1.7&\\
5041.0   & Si\,{\sc ii} &5 & 5041.3   &  42.9 & 31.5&\\
5045.1   &              &  &          &   1.8 &  1.3&\\
5047.8   & He\,{\sc i}  &47& 5047.7   &  19.5 & 14.1&\\
5056.1   & Si\,{\sc ii} &5 & 5056.5   &  24.3 & 17.4&\\
5084.7   & Fe\,{\sc iii}&1F& 5084.8   &   5.7 &  3.9&\\
5132.6   &              &  &          &   1.8 &  1.1&\\
5146.1   & Fe\,{\sc vi} &2F& 5145.8   &   4.9 &  3.0&br.\\
5158.8   & Fe\,{\sc ii}&19F& 5158.8   &   7.4 &  4.4&bl.\\
5169.0   & Fe\,{\sc ii} &42& 5169.0   &   4.9 &  2.9&\\
5176.2   & Fe\,{\sc vi} &2F& 5176.2   &   7.8 &  4.5&br.\\
5191.8   & Ar\,{\sc iii}&  & 5191.8   & $<$0.4&$<$0.2&\\ 
5197.5   & N\,{\sc i}   &1F& 5198.0   &   8.4 &  4.7&\\
         & N\,{\sc i}   &1F& 5200.4   & $<$0.2&$<$0.1&\\
5234.6   & Fe\,{\sc ii} &49& 5234.6   &   8.1 &  4.3&\\
5261.6   & Fe\,{\sc ii}&19F& 5261.6   &   4.3 &  2.2&\\
5270.5   & Fe\,{\sc iii}&1F& 5270.3   &  75.0 & 37.2&\\
5273.4   & Fe\,{\sc ii}&18F& 5273.4   &   3.5 &  1.7&\\
5275.9   & Fe\,{\sc ii} &49& 5276.0   &  10.0 &  4.9&bl.\\
5276.6   & Fe\,{\sc vi} &2F& 5277.8   &       &     &bl.\\
5284.1   & Fe\,{\sc ii} &41& 5284.1   &   4.1 &  2.0&\\
5309.1   & Ca\,{\sc v}  &1F& 5309.2   &  10.4 &  4.8&br.\\
5316.6   & Fe\,{\sc ii} &49& 5316.6   &  24.2 & 11.1&\\
5325.5   & Fe\,{\sc ii} &49& 5325.6   &   3.0 &  1.3&\\
5333.5   &              &  &          &   3.3 &  1.5&\\
5362.9   & Fe\,{\sc ii} &49& 5362.9   &   2.9 &  1.2&\\
5411.6   & He\,{\sc ii} &  & 5411.6   & 101.7 & 39.7&\\
5425.2   & Fe\,{\sc ii} &49& 5425.6   &   4.4 &  1.7&\\
5432.9   & Fe\,{\sc ii}&18F& 5433.1   &   3.8 &  1.4&\\
5460.5   & Ca\,{\sc vi} &2F& 5459.4   &   4.5 &  1.6&\\
5527.4   & Fe\,{\sc ii}&17F& 5527.3   &   2.7 &  0.9&\\
5534.8   & Fe\,{\sc ii} &55& 5534.9   &  14.2 &  4.7&\\
5552.0   &              &  &          &   2.7 &  0.9&\\
5586.2   & Ca\,{\sc vi} &2F& 5586.3   &   3.2 &  1.0&\\
5618.4   & Ca\,{\sc vii}&1F& 5618.6   &  64.9 & 19.4&\\
5666.7   & N\,{\sc ii}  &3 & 5666.6   &  18.4 &  5.2&\\
\hline
\end{tabular}
\end{minipage}
\end{table}

\addtocounter{table}{-1}

\begin{table}
\centering
\begin{minipage}{80mm}
\caption{(continued)}
\begin{tabular}{@{}clrcrrl}
$\lambda_{\rm obs}$&\multicolumn{3}{c}{Identification} &$F_{\lambda}$ &$I_{\lambda}$ &Note\\
(\AA)              & Feature && $\lambda_{\rm air}$     &H$\beta$=100& H$\beta$=100&\\
\hline
5676.1   & N\,{\sc ii}  &3 & 5677.0   &  51.4 & 14.3&\\
5679.6   & N\,{\sc ii}  &3 & 5679.7   &  10.7 &  3.0&\\
5686.2   & N\,{\sc ii}  &3 & 5686.2   &  34.4 &  9.5&\\
5710.8   & N\,{\sc ii}  &3 & 5710.8   &   6.2 &  1.7&\\
5720.8   & Fe\,{\sc vii}&1F& 5721.1   &  58.5 & 15.5&WHT\\
5730.7   & N\,{\sc ii}  &3 & 5730.6   &   2.7 &  0.7&\\
5747.2   & Fe\,{\sc ii}&34F& 5747.0   &   3.0 &  0.8&\\
5754.6  & N\,{\sc ii}  &3F& 5754.6   & 452.0 &114.9&\\
5767.5   &              &  &          &   5.0 &  1.2&\\
5868.4   &              &  &          &   6.6 &  1.5&\\
5876.1   & He\,{\sc i}  &11& 5875.7   &2903.0 &639.9&\\
5920.4   &              &  &          &   2.8 &  0.6&\\
5931.8   & N\,{\sc ii}  &28& 5931.8   &   5.8 &  1.2&\\
5941.7   &              &  &          &   5.9 &  1.2&\\
5953.0   &              &  &          &   4.5 &  0.9&br.\\
5957.0   &Si\,{\sc ii}  &4 & 5957.6   &   1.6 &  0.3&\\
5961.8   &              &  &          &   4.2 &  0.8&\\
5977.0   &              &  &          &   3.5 &  0.7&\\
5979.0   &Si\,{\sc ii}  &4 & 5979.0   &   4.8 &  0.9&\\
6037.7   & He\,{\sc ii} &  & 6036.8   &   4.7 &  0.9&\\
6074.3   & He\,{\sc ii} &  & 6074.3   &   6.4 &  1.1&\\
6086.6   & Fe\,{\sc vii}&1F& 6086.9   & 130.5 & 22.6&\\
         & Ca\,{\sc v}  &1F& 6086.4   &       &     &\\
6170.7   & He\,{\sc ii} &  & 6170.7   &   8.1 &  1.3&\\
6182.3   &              &  &          &  10.4 &  1.6&\\
6233.9   & He\,{\sc ii} &  & 6233.8   &  10.8 &  1.6&\\
6238.3   & Fe\,{\sc ii} &74& 6238.4   &  12.0 &  1.8&bl.\\
6247.5   & Fe\,{\sc ii} &74& 6247.6   &   9.4 &  1.4&bl.\\
6311.3   & S\,{\sc iii} &3F& 6312.0   &  82.0 & 11.1&\\
6317.9   & Fe\,{\sc ii} &--& 6318.0   &   8.4 &  1.1&\\
6347.0   & Si\,{\sc ii} &2 & 6347.5   &  69.5 &  9.1&\\
6371.5   & Si\,{\sc ii} &2 & 6371.3   &  89.7 & 11.4&\\
6383.9   & Fe\,{\sc ii} &--& 6383.5   &  11.0 &  1.4&\\
6402.3   &              &  &          &  10.2 &  1.3&\\
6406.5   & He\.{\sc ii} &  & 6406.5   &  19.7 &  2.4&\\
6416.9   & Fe\,{\sc ii} &74& 6416.9   &  14.3 &  1.7&\\
6432.6   & Fe\,{\sc ii} &40& 6432.6   &  16.1 &  1.9&\\ 
6435.2   & Ar\,{\sc v}  &1F& 6435.1   &  11.3 &  1.3&\\ 
6456.4   & Fe\,{\sc ii} &74& 6456.4   &  36.6 &  4.3&\\
6482.1   & N\,{\sc ii}  &  & 6482.0   &  25.0 &  2.8&\\
6506.4   & Fe\,{\sc ii} &  & 6506.3   &   9.1 &  1.0&\\
6516.0   & Fe\,{\sc ii} &  & 6516.0   &  17.4 &  1.9&\\
6527.2   & He\,{\sc ii} &  & 6527.2   &  18.5 &  2.0&\\
6532.6   &              &  &          &   6.4 &  0.7&\\
6548.1   & N\,{\sc ii}  &1F& 6548.1   & 240.6 & 25.6&WHT\\
6560.6   & He\,{\sc ii} &  & 6560.2   & 484.0 & 51.2&\\
6562.8   & H$\alpha$    &  & 6562.8   &2575.0 &282.0&\\ %
6577.9   &              &  &          &   8.5 &  0.9&\\
6583.5   & N\,{\sc ii}  &1F& 6583.9   & 700.8 & 71.9&\\
6610.5   & N\,{\sc ii}  &  & 6610.5   &   8.5 &  0.9&\\
6678.4   & He\,{\sc i}  &46& 6678.1   &4242.0 &396.8&\\
6683.3   & He\,{\sc ii} &  & 6683.2   &  49.5 &  4.6&\\
6717.0   & S\,{\sc ii}  &2F& 6716.5   &   3.6 &  0.32&bl.\\  %
6730.8   & S\,{\sc ii}  &2F& 6730.7   &   3.1 &  0.27&\\  %
6891.0   & He\,{\sc ii} &  & 6891.0   &  42.2 &  3.2&\\
7005.5   & Ar\,{\sc v}  &1F& 7005.7   &  20.3 &  1.4&br.\\ %
$\ldots$ &              &  &          &       &     &\\
7135.6   & Ar\,{\sc iii}&1F& 7135.8   &  59.2 &  3.6&\\ %
7155.2   & Fe\,{\sc ii} &14F& 7155.2  &  32.3 &  1.9&\\
7170.7   & Ar\,{\sc iv} &2F& 7170.4   &  35.2 &  2.1&br.\\
$\ldots$          &              &  &          &       &     &\\
7236.8   & Ar\,{\sc iv} &2F& 7237.3   &  29.5 &  1.7&\\
\hline
\end{tabular}
\end{minipage}
\end{table}

\addtocounter{table}{-1}

\begin{table}
\centering
\begin{minipage}{80mm}
\caption{(continued)}
\begin{tabular}{@{}clrcrrl}
$\lambda_{\rm obs}$&\multicolumn{3}{c}{Identification} &$F_{\lambda}$ &$I_{\lambda}$ &Note\\
(\AA)              & Feature && $\lambda_{\rm air}$     &H$\beta$=100& H$\beta$=100&\\
\hline
7281.7   & He\,{\sc i}  &45& 7281.3   &2610.0 &141.3&\\
7298.0   & He\,{\sc i}  &  & 7298.0   &  12.8 &  0.7&\\
7320.8   & O\,{\sc ii}  &2F& 7319.7   &   8.5 &  0.5&\\
$\ldots$          &              &  &          &       &     &\\
7377.8   & Ni\,{\sc ii} &2F& 7377.8   &  10.7 &  0.5&\\
7452.5   & Fe\,{\sc ii}&14F& 7452.5   &  13.6 &  0.6&\\
7462.3   &              &  &          &  14.9 &  0.7&\\
7468.3   & N\,{\sc i}   &3 & 7468.3   &  18.8 &  0.9&\\
$\ldots$          &              &  &          &       &     &\\
7593.0   & He\,{\sc ii} &  & 7592.8   &  98.6 &  4.1&\\
7618.3   & N\,{\sc v}?  &  & 7618.5   &  48.9 &  2.0&\\
$\ldots$          &              &  &          &       &     &\\
7711.3   &              &  &          &  68.5 &  2.6&\\
7751.0   & Ar\,{\sc iii}&1F& 7751.1   &  20.0 &  0.7&\\ %
$\ldots$          &              &  &          &       &     &\\
7866.5   &              &  &          &  26.2 &  0.9&\\
7890.0   & Ni\,{\sc iii}&F & 7889.9   &  74.6 &  2.5&\\
7896.3   & Mg\,{\sc ii} &8 & 7896.4   &  29.9 &  1.0&\\
$\ldots$          &              &  &          &       &     &\\
8045.0   & Cl\,{\sc iv} &1F& 8045.6   &   7.4 &  0.2&\\
8084.1   &              &  &          &  27.3 &  0.8&\\
8116.4   &              &  &          &  26.6 &  0.8&\\
$\ldots$          &              &  &          &       &     &\\
8237.0   & He\,{\sc ii} &  & 8236.9   & 236.6 &  6.3&\\
8247.7   &              &  &          &  25.6 &  0.7&\\
8264.9   &              &  &          &  54.2 &  1.4&\\   
$\ldots$          &              &  &          &       &     &\\
8438.0	 & H\,{\sc i}   &18& 8438.0   &  12.6 &  0.29&   \\
8445.6   & O\,{\sc i}   &4 & 8446.4   & 213.7 &  5.0&bl.\\
8467.1   & H\,{\sc i}   &17& 8467.3   &  14.0 &  0.32&\\ %
8480.7   & Cl\,{\sc iii}&3F& 8480.9   &  14.1 &  0.3&\\
8486.2   &              &  &          &  25.3 &  0.6&\\
8499.7   &              &  &          &  98.7 &  2.2&bl.\\
8502.2   & H\,{\sc i}   &16& 8502.6   &  21.3 &  0.48&\\ %
8529.0   &              &  &          &  48.5 &  1.1&\\
$\ldots$          &              &  &          &       &     &\\
8629.2   &              &  &          &  41.9 &  0.9&\\
8648.4   &              &  &          &  62.3 &  1.3&\\
8662.2   &              &  &          &  45.0 &  0.9&\\
8665.0	 & H\,{\sc i}   &13& 8665.0   &  41.0 &  0.84&\\ %
8680.2   & N\,{\sc i}   & 1& 8680.3   & 169.9 &  3.4&\\
8683.5   & N\,{\sc i}   & 1& 8683.4   &  91.3 &  1.8&\\
8686.2   & N\,{\sc i}   & 1& 8686.2   &  61.0 &  1.2&\\
8703.0   & N\,{\sc i}   & 1& 8703.2   &  35.8 &  0.7&\\
8711.5   & N\,{\sc i}   & 1& 8711.7   &  43.9 &  0.9&\\
8719.0   & N\,{\sc i}   & 1& 8718.8   &  33.5 &  0.7&\\
\hline
\end{tabular}
\end{minipage}
\end{table}

\section{Nebular diagnostics and elemental abundances}\label{sect5}

We can now proceed to obtaining estimates of the nebular physical
parameters and abundances. {}From the usual
diagnostic diagram relating H$\alpha$/[S\,{\sc ii}] to [S\,{\sc ii}] 
$\lambda$6717/$\lambda$6731 (see Sabaddin, Minello \& Bianchini 1977) we
find that NaSt1 falls in the photoionization-dominated region.
Therefore the usual nebular diagnostic techniques used for studies of PNe and
symbiotic novae are applicable. 

\subsection{Nebular parameters from observed line fluxes}

In Table~\ref{TABLE2}
we provide observed line fluxes ($F_{\lambda}$) of features visible in
our Keck and WHT observations, including de-reddened fluxes ($I_{\lambda}$),
using the interstellar extinction obtained in Sect.~\ref{sect4}. 
(Table~\ref{TABLE3} contains 
observed and de-reddened IR line fluxes from our UKIRT observations.)
Nebular line identifications have been largely drawn from the lists 
of Kaler et al. (1976), Keyes, Aller \& Feibelman (1990), 
Baluteau et al. (1995), and H.-M. Schmid (priv. comm). 

\begin{figure}
\epsfxsize=7.5cm \epsfbox[102 230 470 590]{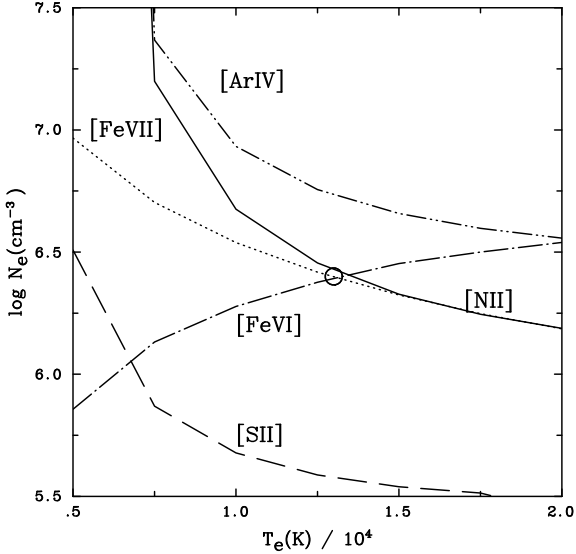}
\caption{Nebular density and temperature diagnostics for NaSt1, including 
N\,{\sc ii} ($\lambda$6583/$\lambda$5755, solid), 
S\,{\sc ii} ($\lambda$6731/$\lambda$4068, dashed), 
Fe\,{\sc vi} ($\lambda$5147/$\lambda$5177, dot-dashed),
Fe\,{\sc vii} ($\lambda$4989/$\lambda$6087, dotted) 
and Ar\,{\sc iv} ($\lambda$4740/$\lambda$7170, dot-dot-dashed).
An open circle denotes the interception point.}
\label{FIG10}
\end{figure}

Despite the rich emission-line spectrum of NaSt1,
many of the usual optical nebular diagnostics are unavailable
(e.g. [O\,{\sc iii}]). Since we do not possess ultraviolet 
(nor are we likely to, given its high reddening!) or far-red 
spectroscopy, we are limited to a small number of available diagnostics. 
A diagnostic diagram for these 
line ratios that are sensitive to $T_{e}$ and $N_{e}$ is presented
in Fig.~\ref{FIG10}. The curves were generated using the {\sc ratio}
program, written by I.D.\,Howarth and S.\,Adams, which solves the
equations of statistical equilibrium allowing a determination of $N_{e}$
as a function of $T_{e}$ for each ratio. 

{}From comparison with studies of symbiotic novae that exhibit a range
of physical conditions (e.g. Schmid \& Schild 1990), we might expect
lines from ions with the highest ionization potentials (IP) such as
[Fe\,{\sc vi-vii}] to sample the highest electron densities, while
lines with low ionization potentials (e.g. [S\,{\sc ii}]) sample lower
density regions. From Fig.~\ref{FIG10}, we see that this may be the
case for NaSt1 since the $N_{e}$ obtained from [S\,{\sc ii}]
$\lambda6731/\lambda4068$ is a factor of ten lower than that obtained
from other diagnostic line ratios.  We note, however, that [S\,{\sc
ii}] $\lambda$4068 is unresolved in our ISIS dataset, and thus the
solution from this ratio should be given a lower weight, since all the
other line ratios are from the HIRES dataset.
Indeed, the solution for [N\,{\sc ii}] is 
consistent with the high ionization potential ions of [Fe\,{\sc vi-vii}].
The interception point of these diagnostics is shown in the figure
(filled circle) at $N_{e}$=3.10$^{6}$cm$^{-3}$ and
$T_{e}$$=$13\,000K.  
The $N_{e}$--$T_{e}$ intersection for the
[Ar\,{\sc iv}] diagnostic ratio (IP$=$60\,eV) with the iron diagnostic
ratios is
at 4.10$^{6}$cm$^{-3}$--20,000K for [Fe\,{\sc vi}] (IP$=$99\,eV)
and 3.10$^{7}$cm$^{-3}$--7,500K for [Fe\,{\sc vii}] (IP$=$125\,eV).

\begin{table}
\centering
\begin{minipage}{80mm}
\caption{Infrared line fluxes based on UKIRT--CGS4 observations.}
\label{TABLE3}
\begin{tabular}{@{}clrlrr}

$\lambda_{\rm obs}$&\multicolumn{3}{c}{Identification} &$F_{\lambda}$ &$I_{\lambda}$ \\
($\mu$m)              & \multicolumn{2}{c}{Feature}
& $\lambda_{\rm air}$     &H$\beta$=100& H$\beta$=100\\
\hline
1.0309  &He\,{\sc i} &3p-6d& 1.0311           &  8.7&  0.08  \\ 
1.0401  &Ni\,{\sc ii}&&   1.0400              &  2.8&  0.03  \\
1.0500  &N\,{\sc i } &&   1.050               &  4.3&  0.04  \\
1.0831  &He\,{\sc i} &2s-2p&1.0830            &2780.0& 21.80  \\ 
1.0914  &Pa$\gamma$  &3-6 &1.0938             & 12.9&  0.10  \\
1.0936  &He\,{\sc ii}&6-12&1.0934             & 12.0&  0.09  \\
1.1011  &He\,{\sc i} &3s-5p &1.1013           &  2.6&  0.02  \\ 
1.1045  &He\,{\sc i} &3p-6d &1.1045           &  2.7&  0.02  \\ 
1.1125  &Fe\,{\sc ii}&& 1.113                 &  2.7&  0.02  \\
1.1292  &Fe\,{\sc vi}&& 1.13                  &  2.2&  0.02  \\
$\ldots$&            &&                       &     &        \\
1.2469  &N\,{\sc i}  && 1.2469                &  2.2&  0.01  \\
1.2527  &He\,{\sc i} &3s-4p &1.2526           & 43.9&  0.21  \\ 
1.2792  &He\,{\sc i} &3d-5f &1.2785           & 52.6& 0.24   \\ 
1.2823  &Pa$\beta$   &3-5&   1.2818           & 34.4& 0.15   \\
1.2979  &He\,{\sc i} &3d-5p &1.2985           & 14.9& 0.06   \\ 
$\ldots$&            &&                       &     &        \\
1.6400  &H\,{\sc i}  &12-4& 1.6412            & 7.6 & 0.02   \\
1.6796  &H\,{\sc i}  &11-4& 1.6811            & 5.7 & 0.01   \\
1.7008  &He\,{\sc i} &3p-4d&1.7002            &101.5& 0.24   \\ 
1.7351  &H\,{\sc i}  &10-4& 1.7367            & 10.0& 0.02   \\
$\ldots$&            &&                       &     &        \\
2.0594  &He\,{\sc i} &2s-2p &2.0581           &1044.0& 1.82  \\ 
2.1132  &He\,{\sc i} &3p-4s &2.1120           & 34.7&  0.06  \\ 
2.1642  &Br$\gamma$  &4-7 &2.1655             & 20.3&  0.03  \\
2.1889  &He\,{\sc ii}&7-10&2.1889             &  2.3&$<$0.01 \\
\hline
\end{tabular}
\end{minipage}
\end{table}

\subsection{Nebular abundances}\label{5.2}

We now utilise our derived nebular parameters to obtain estimates of
elemental abundances, and consider first the abundances of the
collisionally-excited species.  Ionic abundances were obtained by
solving the equations of statistical equilibrium using the {\sc equib}
program, also written by S.\,Adams and I.D.\,Howarth,
and are given in Table~\ref{TABLE4}. Total element abundances
were derived using the ionization correction factor (ICF) scheme 
of Kingsburgh \& Barlow (1994) and these are also listed in Table~\ref{TABLE4}.

\begin{table}
\centering
\begin{minipage}{80mm}
\caption{Metal ionic and total abundances for NaSt1 relative to hydrogen, 
in units of 10$^{-6}$, including ionization correction factors
(ICFs), assuming $T_{e}$=13\,000K, $N_{e}$=3.10$^{6}$cm$^{-3}$.
Lines used in the analysis are shown in parenthesis.}
\label{TABLE4}
\begin{tabular}{@{}l@{\hspace{3mm}}r@{\hspace{3mm}}r@{\hspace{3mm}}r@{\hspace{3mm}}r@{\hspace{3mm}}
r@{\hspace{3mm}}r}
Element & 
$\frac{{\rm X}^{+}}{{\rm H}^{+}}$ &
$\frac{{\rm X}^{2+}}{{\rm H}^{+}}$ &
$\frac{{\rm X}^{3+}}{{\rm H}^{+}}$ &
$\frac{{\rm X}^{4+}}{{\rm H}^{+}}$ &
ICF(X) & $\frac{\rm X}{\rm H}$ \\
\hline
N  & 256.00 &-- & -- & -- & 2.90 & 743.00 \\
   &(6584) \\
O & 1.25 & 2.38 & -- & -- & 1.00 & 3.63 \\
       &(7320)&(4959) \\
Ne & --  & 51.00& -- & -- & 1.52 & 77.80 \\
        && (3869) \\
S & 2.53& 5.34 & --   & --   & 1.12 & 8.80 \\
        &(6731)&(6312)& \\
Ar &  -- & 0.28 & 1.33 & 0.15 & 1.52 & 2.69 \\
        &     &(7135)&(4740)&(7005)& \\
\hline
\end{tabular}
\end{minipage}
\end{table}

Turning to the helium abundance,
optical He\,{\sc i-ii} transitions are produced by
radiative recombination, with additional
contributions to He\,{\sc i} line strengths from
collisional processes. Helium ionic
abundances were derived relative to H$^{+}$ using Case~B recombination theory.
For this we utilised interpolated effective recombination coefficients
from Osterbrock (1989) for H\,{\sc i} and He\,{\sc ii}, and He\,{\sc
i} coefficients from Smits (1996). He\,{\sc i} line strengths have
been corrected for the effect of collisional population of their upper
states (Kingdon \& Ferland 1995).  These collisional factors (C/R) are
shown in Table~\ref{TABLE5} for $N_{e}$=3.10$^{6}$cm$^{-3}$ and
$T_{e}$=13\,000\,K together with the derived helium ionic abundance
ratios.  It is clear that there is a large scatter in the He$^+$/H$^+$
ratios which must result from optical depth effects.
Indeed, the two lowest-lying singlets at $\lambda6678$ and
$\lambda7281$ show the highest ionic abundances. We cannot therefore
derive a reliable He$^+$/H$^+$ ratio using Case~B recombination
theory. The He$^{2+}$/H$^+$ ratio is, however, better determined since
the He\,{\sc ii} $\lambda$4686 and $\lambda$5412 transitions are in
the correct Case~B ratios and yield a mean He$^{2+}$/H$^+$ ratio of
0.64. Furthermore, with this abundance, the predicted strengths of
He\,{\sc ii} $\lambda$4860 and $\lambda$6683 are 28.0 and 3.8 relative
to H$\beta$, in agreement with the measured de-reddened values of
34.9 and 4.6. 

In summary, while we cannot determine a reliable He$^+$/H$^+$ abundance
because the lines are optically thick, we can determine 
a lower limit to the He/H abundance by using the He$^{2+}$/H$^+$ ratio of 0.64.
In reality, we expect the total He/H abundance to be at least twice this,
given the strength of the He\,{\sc i} lines and the expectation that
He$^+$ is the dominant ionization stage of He. Even with this lower limit,
it is apparent that NaSt1 is extremely helium-rich.

\begin{table}
\centering
\begin{minipage}{80mm}
\caption{Helium-to-hydrogen ionic abundance ratios and collisional
factors (C/R) for NaSt1 
at $T_{e}$=13\,000K, $N_{e}$=3.10$^{6}$cm$^{-3}$.}
\label{TABLE5}
\begin{tabular}{@{}rrrrr}
Line & $\frac{I \times 100}{I_{H\beta}}$ & C/R 
& $\frac{N({\rm He}^{+})}{N({\rm H}^{+})}$ 
& $\frac{N({\rm He}^{2+})}{N({\rm H}^{+})}$ \\
\hline
He\,{\sc i} $\lambda$3889 & 251.5 & 0.51 & 2.04 & --\\
He\,{\sc i} $\lambda$4387 &  31.8 & 0.09 & 6.70 & --\\
He\,{\sc i} $\lambda$4471 &  78.7 & 0.22 & 1.83 & --\\
He\,{\sc i} $\lambda$4922 &  59.9 & 0.11 & 5.78 & --\\
He\,{\sc i} $\lambda$5876 & 639.9 & 0.44 & 4.77 & -- \\
He\,{\sc i} $\lambda$6678 & 396.8 & 0.17 & 12.85 & -- \\
He\,{\sc i} $\lambda$7281 & 141.3 & 0.83 & 14.46 & -- \\
He\,{\sc ii} $\lambda$4686& 546.5 & --   & --  & 0.67\\
He\,{\sc ii} $\lambda$5412&  39.7 & --   & --  & 0.61\\
\noalign{\smallskip}
\hline
\end{tabular}
\end{minipage}
\end{table}

In Table~\ref{TABLE6} we present a summary of the abundances derived
for NaSt1 and a comparison with other objects. First, we find that the
abundances of Ne, Ar and S are very similar to the average H\,{\sc ii}
region values (Shaver et al. 1983). This excellent agreement suggests
that the use of single representative $T_e$ and $N_e$ values has
produced reliable abundances. The N/O ratio for NaSt1 is very
different to that expected for H\,{\sc ii} regions because N is
enhanced by a factor of 20 while O is depleted by a factor of 140.
Such extreme values indicate heavily CNO-processed material. Indeed,
the total H\,{\sc ii} C$+$N$+$O abundance of $8.27 \times 10^{-4}$ is
comparable to the combined N$+$O abundance of $7.47 \times 10^{-4}$
for NaSt1 suggesting that very nearly all the carbon and oxygen have
been processed to nitrogen. This is in accord with the null detection
of carbon emission lines in NaSt1 (Sect. 3.1).

The only object known to us which shows such extreme CNO-processing is
the LBV $\eta$ Car. In Table~\ref{TABLE6}, we list the abundances from
the recent study of Dufour et al. (1997). The spectacular bipolar
nebula associated with $\eta$ Car was ejected in 1840 during a giant
eruption (Davidson \& Humphreys 1997). The abundances show that
the ejected outer stellar layers are composed of CNO-equilibrium
products. Other LBV nebulae (e.g. AG Car, R127, S119) have much
smaller nitrogen enrichments of 4--11, and little, if any, oxygen
depletion, indicative of CN-processing only (Smith et al. 1997, 1998).
These abundance differences have led Lamers et al. (1998) to propose
that the LBV-like star whose spectrum now dominates the nucleus was
not the star that ejected the nebula because the appearance of
the stellar spectrum is indicative of mildly-enhanced CN products.
They instead suggest that the eruptor was more evolved (possibly
WR-like) and thus more massive than the LBV $\eta$ Car.

The abundances we determine for NaSt1 are also very different to those
measured for symbiotic stars. Nussbaumer et al. (1988) and Schmid \&
Schild (1990) find CNO abundances similar to those observed in red
giants with only N enhanced from CN-processing. The abundances
we determine are also not in agreement with those measured for novae
which show enriched C, N and O and sometimes Ne (Livio \& Truran 1994).

\begin{table}
\centering
\begin{minipage}{80mm}
\caption{Summary and comparison of abundances for NaSt1.
He/H and N/O are by number, while those of other elements 
are 12+log X/H.
H\,{\sc ii} region abundances are from
Shaver et al. (1983), while those of $\eta$ Car are taken from 
Dufour et al. (1997).}
\label{TABLE6}
\begin{tabular}{@{}lrrr}
Abundance      &NaSt1     &$\langle$H\,{\sc ii}$\rangle$
& $\eta$ Car \\
\hline         
He/H           & $>0.64$  & 0.10 & 0.18 \\
N/O            & 200      & 0.07 & $>56$ \\
\noalign{\smallskip}
12+log(O/H)            & 6.56     & 8.70 & $<7.29$ \\
12+log(N/H)           & 8.87     & 7.57 & 9.04  \\
12+log(Ne/H)           & 7.89     & 7.90 & 7.98 \\
12+log(Ar/H)          & 6.43     & 6.42 \\
12+log(S/H)           & 6.94     & 7.06 \\
\hline
\end{tabular}
\end{minipage}
\end{table}

\begin{figure*}
\epsfxsize=17.6cm \epsfbox[34 250 516 575]{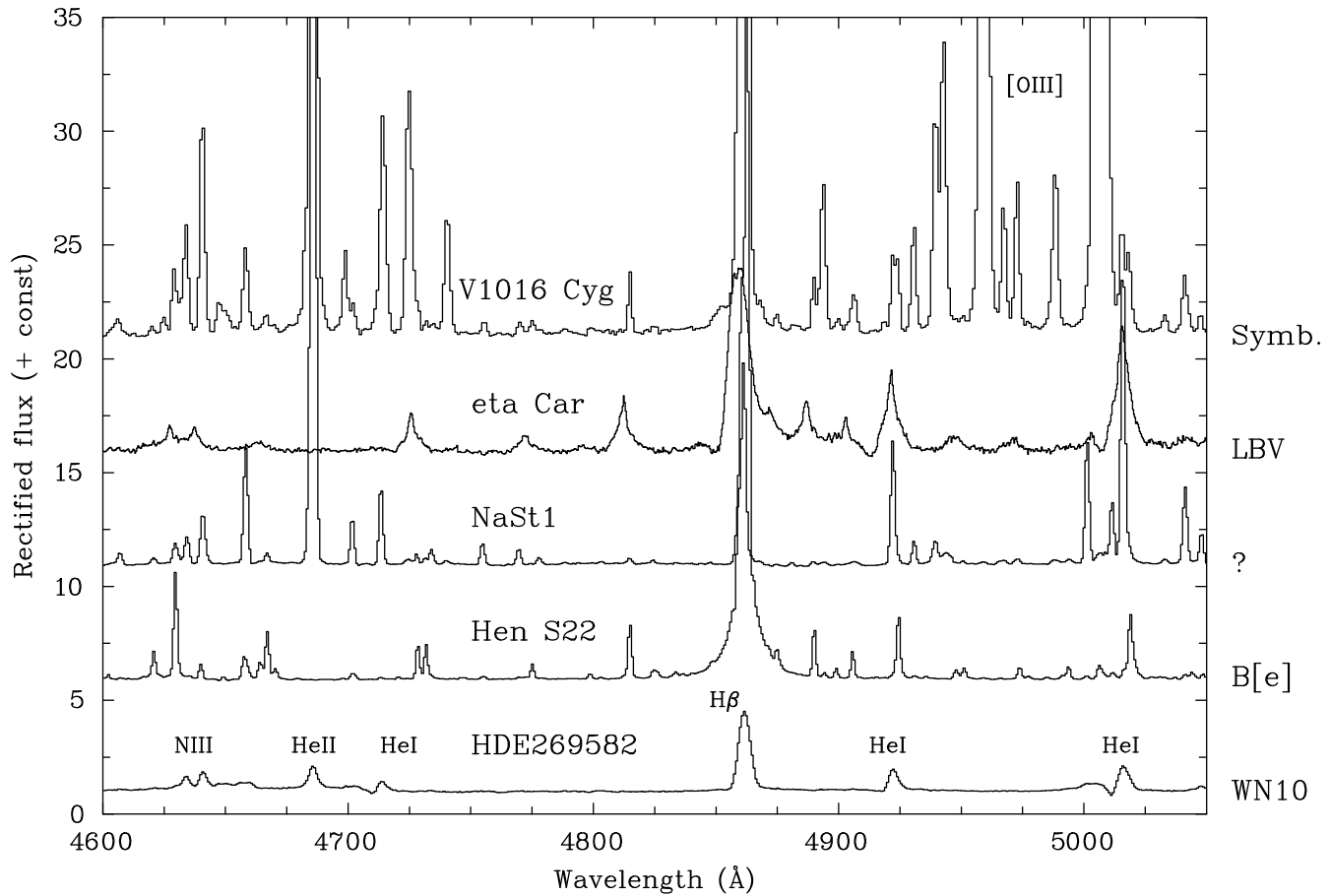}
\caption{Comparison between the 
intermediate dispersion $\lambda\lambda$4600--5050 spectral region
of NaSt1 and 
various emission line objects:
the LMC WN10 star 
(previously Ofpe/WN9) HDE\,269582; the LBV $\eta$ Car; the LMC B[e] star 
Hen~S22; and the Galactic D-type symbiotic
nova V1016 Cyg (Schmid \& Schild 1990). Observations are 
from Crowther \& Smith (1997), B. Bohannan (priv. comm.), 
our own AAT-RGO spectrograph observations from 1994 December,
and H.-M. Schmid (priv. comm.), respectively}
\label{FIG11}
\end{figure*}                

\section{Discussion}\label{sect6}

In this final section, we discuss the information that we 
have obtained for NaSt1 with the aim of identifying 
its true nature.

\subsection{Comparison of NaSt1 with other peculiar emission line objects}\label{comparison}

First, we will compare our observations of NaSt1 with known 
WR, Ofpe/WN9, LBV and B[e] stars, to which it has
previously been compared, and symbiotic novae, to which
it shows certain similarities.

\subsubsection{Wolf-Rayet and Ofpe/WN9 stars}

A Wolf-Rayet (WN10) classification for NaSt1 was suggested by
Nassau \& Stephenson (1963) and supported by Massey  
\& Conti (1983), while van der Hucht et al. (1989) proposed an
Ofpe/WN9 classification. WR stars represent the  final 
state in the evolution of very massive 
stars prior to  the supernova explosion. Their optical spectrum is 
characterised  by broad, pure emission  (and P Cygni) profiles of 
highly excited species resulting from a  fast, extremely dense stellar wind. 
Ofpe/WN9 stars, reclassified as WN9--11 stars by Smith et al. (1994)
and Crowther, Hillier \& Smith (1995a), are intimately related to 
both classical WR stars and LBVs.

In Fig.~\ref{FIG11}, we compare our rectified WHT-ISIS spectrum of
NaSt1 with the LMC star HDE\,269582 (WN10, previously Ofpe/WN9;
Crowther \& Smith 1997). This confirms the quite different appearance of 
NaSt1, even at intermediate spectral resolution. Stellar 
He\,{\sc i--ii}, H\,{\sc i} and N\,{\sc iii}
features are observed in the WN10 star, including the P Cygni He\,{\sc i}
$\lambda$5016 profile, providing evidence for a dense stellar wind outflow.
The only nebular lines observed in WR 
stars are those weak features originating in low-excitation, 
ejecta-type circumstellar nebulae (e.g. Nota et al. 1996).

\subsubsection{B[e] stars}

van der Hucht et al. (1989) suggested a possible B[e] supergiant
nature for NaSt1 -- such luminous, massive objects appear to 
have a equatorial excretion disk plus a polar OB-type stellar wind. 
In the optical, their spectra are characterised by a plethora of
low-excitation emission lines including  the Balmer series plus 
narrow permitted and forbidden lines of  singly ionized ions 
(e.g. Fe\,{\sc ii}), as shown in Fig.~\ref{FIG11} for the very 
luminous LMC B[e] star Hen~S22. The presence of 
He\,{\sc ii} $\lambda$4686 emission in NaSt1 led van  der Hucht 
et al. (1989, 1997) to tentatively suggest that it may be a 
high temperature counterpart of B[e] stars and hence they proposed a 
new O[e] classification.

In common with B[e] stars, NaSt1
shows significant mid-IR dust excess and [Fe\,{\sc ii}] lines 
in its optical spectrum. However, NaSt1 additionally  displays 
high excitation, permitted (He\,{\sc ii} $\lambda$4686) and 
forbidden ([Fe\,{\sc vii}] $\lambda$6087) emission lines and does not 
show the broad, stellar Balmer lines (compare H$\beta$ profiles
in Fig.~\ref{FIG11}). 

\begin{figure}
\epsfxsize=8.5cm \epsfbox[14 160 462 540]{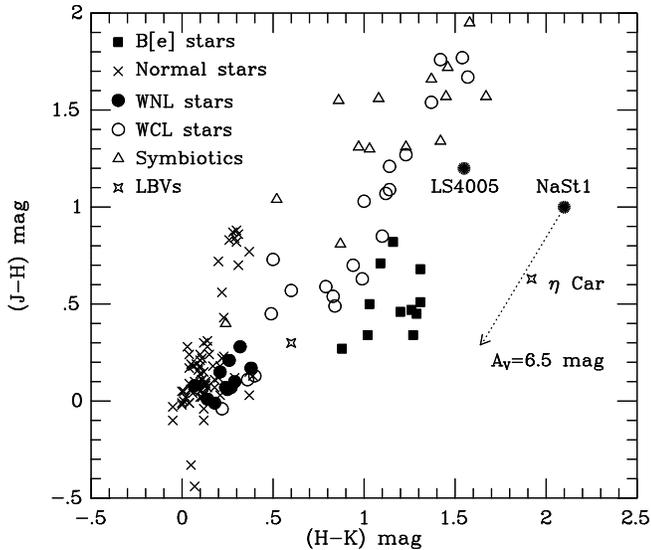}
\caption{Two--colour IR diagram (J$-$H versus H$-$K) for
various luminous objects -- normal stars and supergiants,
B[e] supergiants, WN6--11 stars, LBVs, `dusty' WCL stars 
and D-type symbiotic novae. Galactic objects are shows as
open symbols while LMC stars are filled-in. A correction
for the high interstellar reddening towards NaSt1 is shown, while
other Galactic sources may require comparable correction
(particularly WCL stars and symbiotics). Observational data
is taken from Gummersbach, Zickgraf \& Wolf (1995), McGregor, Hillier \& Hyland
(1988), Williams et al. (1987), Munari et al. (1992) and
Crowther \& Smith (1997)}
\label{FIG12}
\end{figure}                

\subsubsection{Symbiotic novae}

The spectral appearance of NaSt1 shares some 
characteristics with symbiotic novae. These systems consist 
of a red giant, the ionized nebula and a hot ionizing source, 
with the red star generally directly observable in the near-infrared. 
Fig.~\ref{FIG11} presents a comparison with
the D-type (`dusty') symbiotic nova V1016 Cyg, which Schmid \& Schild (1990)
found to have an electron density of $N_{e}$$\sim$10$^{6}$cm$^{-3}$,
comparable to NaSt1. 

V1016 Cyg and NaSt1 show a strong nebular spectrum, including 
He\,{\sc ii} $\lambda$4686, H$\beta$, [N\,{\sc ii}], N\,{\sc iii},
plus low and high-excitation forbidden Fe emission. Both systems 
show IR dust emission and a correlation between density and 
ionization potential (Schmid \& Schild  1990). 
However, significant differences are also found:
(i) there is no spectroscopic evidence for a red giant in NaSt1
at IR wavelengths; (ii) the He\,{\sc i-ii} emission spectrum of 
NaSt1 is dramatically stronger;
(iii) the characteristic symbiotic  O\,{\sc vi} Raman scattered lines are 
absent; (iv) [O\,{\sc iii}] $\lambda$5007 emission 
is extremely strong in V1016 Cyg, and indeed all dusty symbiotics,
which is very weak in NaSt1. Therefore, while NaSt1 shows some 
similarities with dusty symbiotic novae, its nebular line strengths are 
anomalous, and its overall properties are distinct.

\subsubsection{$\eta$ Carinae}

From Sect.~\ref{5.2}, the abundance pattern of NaSt1 most closely
resembles $\eta$ Car. Consequently we include a spectroscopic comparison 
between NaSt1 and $\eta$ Car in Fig.~\ref{FIG11}. $\eta$ Car shows narrow emission
lines of H\,{\sc i} and He\,{\sc i} with very broad wings, and He\,{\sc
ii} $\lambda$4686 is absent.  Highly ionized forbidden nebular lines
are not observed in $\eta$ Car. Therefore, although its appearance 
is also unusual, $\eta$ Car bears little spectroscopic resemblance to NaSt1. 

\subsection{What is the nature of NaSt1?}

{}From the objects discussed in the previous subsection, NaSt1 most
closely resembles symbiotic novae from optical spectroscopy and $\eta$
Car from nebular abundances.  We believe we can rule
out a symbiotic nature on the basis of its He-rich, O-deficient
chemistry and the spectroscopic absence of a red-giant component.
Specifically, the CNO-cycle products in the ejected nebula are unique
to a massive post-main sequence star. We also note that 
NaSt1 does not appear to be a strong X-ray emitter. Pollock (1987) suggested
that NaSt1 was a possible X-ray source from {\it Einstein} data.
NaSt1, however, lies fairly close (20$'$) to the well-known supernova 
remnant Kes~79 (G33.6+0.1). Higher quality {\it ROSAT} PSPC 
observations revealed that the majority of the X-ray 
emission near NaSt1 is associated with  Kes~79 (Pollock, priv. comm.) 
and negligible emission is observed from NaSt1 itself.

The unusual nature of NaSt1 is further illustrated in 
Fig.~\ref{FIG12} which compares its two--colour IR 
index (J$-$H and H$-$K) with other (potentially related) objects
from our Galaxy (open symbols) and the LMC (filled-in).
Included are `normal' stars, B[e] stars, supergiants,  WN6--11
stars, LBVs, `dusty' 
late WC (WCL) stars, and D-type symbiotic  novae. We also show the 
location of NaSt1 before and after correction for the 
high interstellar extinction. We find the 
IR characteristics of NaSt1 are extremely unusual. Most other
dusty luminous objects, including B[e], WCL and symbiotics, show quite 
different IR characteristics. The only object known to 
possess similar IR properties, after correction for the interstellar 
reddening towards NaSt1, is $\eta$ Car. 

\subsection{A massive, evolved star cloaked by a dense ejecta nebula}

Despite the absence of a stellar signature in NaSt1, our nebular
analysis suggests that it contains a hot, luminous, evolved star,
hidden from direct view by the dense nebular envelope.  The
composition of the nebula indicates that the star ejected its outer
layers when CNO-equilibrium products were present on the surface.
Comparison with the lower limit on the He/H ratio we derive of 0.64
and the surface He/H ratios derived for WR stars (Crowther et al.
1995a, Crowther \& Smith 1997) suggests that the central star must have
been a WN star at the time of eruption rather than an LBV-type star.

An early WN (WNE) star identification agrees with the high temperature we
find for the ionizing source. These 
stars are highly evolved objects, with stellar temperatures greatly 
in excess of 30,000\,K and hydrogen-deficient, CNO-processed stellar winds. 
Indeed, our estimate of the stellar luminosity lies in the 
range occupied by early WN stars -- log~($L_/L_{\odot}$)=5.1--6.1
(Crowther, Smith \& Hillier 1995b; Hamann \& Koesterke 1998). 
Another possibility is that the remnant star would now have a more
advanced chemical composition than a WN star if most of the stellar
envelope was lost in the ejection, namely a WC-type star, also in 
accord with the temperature and luminosity we derive for the ionizing 
source. 

Early-type WR stars, however, have very dense stellar winds, so how would the
characteristic broad, stellar features not be directly observed? 
Perhaps its stellar
wind has not yet pierced the dense ejected nebula.  The dynamical age
obtained from the nebular analysis is of the order of a few thousand
years. This would imply a very recent ejection of a large amount of
material, as evidenced from the very high characteristic electron
density obtained of $3 \times 10^6$\,cm$^{-3}$. The only spectral
feature which shows any possible type of stellar wind outflow is the
He\,{\sc i} $\lambda$20581 profile which has wings extending to
$\sim$300~km\,s$^{-1}$, and closely resembles the He\,{\sc i} 
1.0830$\mu$m profile of the massive young stellar object (YSO) 
Sh~2-106IR, with a comparable outflow velocity (Drew, Bunn 
\& Hoare 1993 and J.~Drew, priv. comm.). 
This velocity is, however, much lower than
early WR stars which have winds of $\sim$2000--3000\,km\,s$^{-1}$.  
This velocity is more characteristic of an LBV during its hot phase
(e.g. AG Car; Smith, et al. 1994). Theoretically, we might reconcile
a high stellar temperature and low wind velocity with a massive star that 
is extremely close to its Eddington limit. Indeed, low velocity, aspherical 
outflows are anticipated for stars close to the related `$\Omega$ limit' 
which includes the effect of rotation (see Langer 1997). 

The only LBV known which has an ejected nebula composed of heavily
processed CNO material is $\eta$ Car. As discussed in Sect.~\ref{5.2}, 
the advanced evolutionary state of this nebula has led Lamers et al.
(1998) to propose that the eruptor was not the LBV but an unseen more
evolved star.  The IR two--colour index of NaSt1 in Fig.~\ref{FIG12}
shows strong similarities with $\eta$ Carinae, indicating comparable
nebular dust conditions. It is possible that 
NaSt1 is a counterpart to $\eta$ Carinae with an unseen 
massive evolved central star that underwent a major instability and
ejected its outer layers a few thousand years ago. Differences in the
spectral appearances of NaSt1 and $\eta$ Car are probably attributable 
to geometry, age, and the hotter ionizing source in NaSt1 ($\eta$ Car
is cooler than $\sim$30kK from the absence of He\,{\sc ii} 
$\lambda$4686 emission).  
Unfortunately, we are unable  to comment on details of the precise geometry of 
NaSt1 since we do not possess deep, high spatial resolution optical/IR 
imaging.

Whatever the true nature of NaSt1, its properties are extremely unusual.
Is there any evidence for objects with similar characteristics?
van der Hucht et al. (1984) and Williams et al. (1987) have discussed
similarities between NaSt1 and LS4005 (WR85a). LS4005 also shows narrow 
($\Delta\lambda\sim 20$\,km\,s$^{-1}$) emission lines of N\,{\sc ii-iii},
Fe\,{\sc ii-iii} (both allowed and forbidden), with strong He\,{\sc i-ii}
features, and no absorption features present, and photometric
variability (van der Hucht et al. 1989). LS4005 would certainly represent 
an excellent target for future high resolution spectroscopy and imaging.

\subsection{Summary and future work}

We have presented optical and IR spectroscopy, and imaging of NaSt1
which have revealed a heavily CNO-processed nebula.  NaSt1 serves as a
useful reminder that great care should be taken when selecting IR
sources to be classification standards. While many authors have
commented that NaSt1 bears little resemblance to other late-WN type
stars at IR wavelengths, it has nevertheless remained as a
classification standard.

We interpret the spectrum of NaSt1 as arising in a dense nebula, 
ejected by an evolved massive star. 
The H-deficient, CNO-processed nebula suggests that an unseen early 
WN or WC star provides the ionizing flux. The
only object which shares some of the peculiar characteristics of
NaSt1 is $\eta$ Carinae.
NaSt1 appears to be a remarkable
object, and hints at new insights into massive star evolution. 

\section*{Acknowledgments}
We would like to thank Mike Barlow,  
Steve Fossey, Jay Gallagher, Norbert Langer, Mario Livio, 
Xiao Wei Lui, Andy Pollock and Hans-Martin Schmid for many fruitful 
discussions. We also wish to thank Bruce 
Bohannan, You-Hua Chu, Hans-Martin Schmid,
Lindsey Smith and Peter Tamblyn for generously forwarding additional
observations.  PAC acknowledges financial support from PPARC and the 
Royal Society.

We are especially grateful to Marten van Kerkwijk for obtaining the Keck 
HIRES observations for us, and the staff of the now defunct Royal Greenwich 
Observatory for  obtaining service spectroscopy and imaging. The
William Herschel  Telescope is operated on the  Island of La Palma by
the  Isaac Newton Group in the Spanish  Observatorio del Roque de los
Muchachos of the Instituto de Astrofisica de  Canarias. The W.M.~Keck
Telescope is operated by Caltech and the University of California on
Mauna Kea, Hawaii, while the  U.K. Infrared Telescope is operated by
the Joint Astronomy Centre on  behalf of the Particle Physics and
Astronomy Research Council also on Mauna Kea, Hawaii.

\bsp

\label{lastpage}


\begin{thebibliography}{}
\bibitem{X} Baluteau J.-P., Zavagno A., Morisset C., P\'{e}quignot D., 
           1995, A\&A 303, 175
\bibitem{X} Blum R.D., DePoy D.L., Sellgren K., 1995, ApJ 441, 603
\bibitem{X} Brand J., Blitz, L., 1993, A\&A, 275, 67
\bibitem{X} Cappellaro E., Benetti S., Sabaddin F., Salvadori L., Torrato M., Zanin C., 1994, MNRAS, 267, 871
\bibitem{X} Crowther P.A.,  Smith L.J., 1997, A\&A 320, 500
\bibitem{X} Crowther P.A., Hillier D.J., Smith L.J., 1995a, A\&A 293, 172 
\bibitem{X} Crowther P.A., Smith L.J., Hillier D.J., 1995b, A\&A 302, 457
\bibitem{X} Daly P.N., Beard S.M., 1992, Starlink User Note 27 (Rutherford
            Appleton Laboratory)
\bibitem{X} Dame T.M., Thaddeus P., 1985, ApJ 297, 751
\bibitem{X} Davidson K., Humphreys R.M., 1997, ARAA, 35, 1
\bibitem{} Drew J.E., Bunn J.C., Hoare M.G., 1993, MNRAS 265, 12
\bibitem{X} Dufour, R.J., Glover, T.W., Hester, J.J., Currie, D.G., van
            Orsow, D., Walter, D.K., 1997, in Luminous Blue Variables: 
            Massive Stars in Transition, ed. A. Nota \& H.J.G.L.M. Lamers 
            (ASP Conf. Ser.), 120, p. 255
\bibitem{} Figer D.F., McLean I.S., Najarro F., 1997, ApJ 486, 420
\bibitem{X} Gummersbach C.A., Zickgraf F.-J., Wolf B., 1995, A\&A 302, 409
\bibitem{X} Hamann W.-R., Koesterke L., 1998, A\&A, 333, 251
\bibitem{X} Henize K.G., 1967, ApJS 14, 125
\bibitem{X} Henize K.G., 1976, ApJS 30, 491
\bibitem{X} Herbig G.H., 1995, ARA\&A 33, 19
\bibitem{X} Horne K., 1986, PASP 98, 609
\bibitem{X} Howarth I.D., Murray J., Mills D., Berry D.S., 1995,
           Starlink User Note, 50.16 (Rutherford Appleton Laboratory)
\bibitem{X} van~der Hucht K.A., Conti P.S., Lundstr\"{o}m I., Stenholm B., 
           1981, Space Sci. Rev. 28, 227
\bibitem{X} van der Hucht K.A., Williams P.M., Th\'{e} P.S., 1984, in:
           Observational Tests of the Stellar Evolution Theory, IAU
           Symp. 105, A. Maeder \& A. Renzini (eds.), Kluwer p.~273
\bibitem{X} van der Hucht K.A., Williams P.M., van Genderen A.M.,
           Mulder P., Zealey W.J., 1989, in:
           Physics of Luminous Blue Variables, Davidson, K., Moffat, A.F.J., Lamers, 
           H.J.G.L.M., (eds)., Kluwer, Dordrecht, p.~301
\bibitem{X} van der Hucht K.A., Williams P.M., Moris P.M., van Genderen A.M.,
           1997, in: Luminous Blue Variables: Massive Stars in Transition, 
           Nota, A., Lamers H.J.G.L.M. (eds)., ASP Conf. Series 120, p.211
\bibitem{X} Kaler J.B., Aller L.H., Epps, H.W., Czyzak S.J., 
           1976, ApJS 31, 163
\bibitem{X} Keyes C.D., Aller L.H., Feibelman W.A., 1990, PASP 102, 59
\bibitem{X} Kingdon J., Ferland G.J., 1995, ApJ 442, 714
\bibitem{X} Kingsburgh R.L., Barlow M.J,, 1994, MNRAS 271, 257
\bibitem{X} Lamers H.J.G.L.M., Livio M., Panagia N, Walborn N.A.,
            1998, ApJ, 505, L131
\bibitem{X} Langer N., 1997, in Nota A., Lamers, H.J.G.L.M. eds,
           Luminous Blue Variables, Massive Stars in Transition,
           ASP Conf. Series, Vol 120, p.~83
\bibitem{X} Le Bertre T., Lequeux J., 1993, A\&A 274, 909
\bibitem{X} Livio M., Truran J.W., 1994, ApJ, 425, 797
\bibitem{X} Massey P., Conti P.S., 1983, PASP 95, 440
\bibitem{X} McGregor P.J., Hillier D.J, Hyland A.R., 1988,
            ApJ, 334, 639
\bibitem{X} Miller G.J., Chu Y.-H., 1993, ApJS 85, 137
\bibitem{X} Mills D., Webb J., 1994, Rutherford Appleton Laboratory, SUN 152.1
\bibitem{X} Morris P.W., Eenens P.R.J., Hanson M.M., Conti P.S., Blum R.D.,
           1996, ApJ, 470, 597
\bibitem{X} Munari U., Iudin B.F., Taranova O.G., Massone G., Marang F.,
Roberts G., Winkler H., Whitelock P.A., 1992, A\&AS 93, 383
\bibitem{X} Nassau J.J., Stephenson C.B., 1963, Luminous Stars in the Northern
           Milky Way, IV, Hamburg-Bergedorf
\bibitem{X} Nussbaumer H., Schild H., Schmid H.-M., Vogel M., 1988,
            A\&A, 198, 179
\bibitem{} Nota A., Pasquali A., Drissen L., Leitherer C., Robert C., Moffat A.F.J., Schmutz W., 1996, ApJS 102, 383
\bibitem{X} Pollock A.M.T., 1987, ApJ 320, 283
\bibitem{X} Osterbrock D.E., 1989, Astrophysics of Gaseous Nebulae \& Active
           Galactic Nuclei, University Science Books
\bibitem{X} Sabaddin F., Minello S., Bianchini A., 1977, A\&A 60, 147
\bibitem{X} Schmid, H.-M., Schild H., 1990, MNRAS 246. 84
\bibitem{X} Shaver P.A., McGee R.X., Newton L.M., Danks A.C., Pottasch
            S.R., 1983, MNRAS, 204, 53
\bibitem{X} Shortridge, K., Meyerdierks, H., Currie, M., Clayton, M.,
           1997, Starlink User Note 86.13, Rutherford Appleton Laboratory
\bibitem{x} Shylaja B.S., Anandarao B.G., 1993, Astro. Space Sci. 205, 283
\bibitem{X} Smith L.F., 1968, MNRAS 140, 409
\bibitem{X} Smith L.J., Crowther P.A., Prinja, R.K., 1994, A\&A, 281, 833
\bibitem{X} Smith L.J., Stroud M.P., Esteban C., V\'{i}lchez, J.M., 1997, MNRAS 290, 265
\bibitem{X} Smith L.J., Nota A., Pasquali A., Leitherer C., Clampin
            M., Crowther P., 1998, ApJ, 503, 278
\bibitem{X} Smits D.P., 1996, MNRAS 278, 683
\bibitem{} Storey P.J., Hummer D.G., 1995, MNRAS 272, 41
\bibitem{X} Tamblyn P., Rieke G.H., Hanson M.M., Close L.M., McCarthy D.W. Jr, Rieke M.J., 1996, ApJ 456 206
\bibitem{X} Williams P.M., van der Hucht K.A., Th\'{e} P.S., 1987, A\&A 182, 91
\end{thebibliography}
\end{document}